\documentclass[Journal]{IEEEtran}
\IEEEoverridecommandlockouts
\usepackage{cite}
\usepackage{notoccite}
\usepackage{float}
\usepackage{amsmath,amssymb,amsfonts}
\usepackage[ruled,linesnumbered,vlined]{algorithm2e}
\usepackage{graphicx}
\usepackage{subfigure} 
\usepackage{subcaption}
\usepackage{textcomp}
\usepackage{xcolor}
\usepackage{multirow}
\usepackage{tabularx}
\usepackage{diagbox}
\usepackage{colortbl}
\usepackage{breakurl}        
\usepackage{array}
\usepackage{booktabs}   
\usepackage{footnote}
\usepackage{makecell} 
\makesavenoteenv{tabular} 
\makesavenoteenv{table}
\usepackage{subfigure}
\usepackage{siunitx}    
\usepackage{lettrine}
\usepackage{threeparttable}
\usepackage{tabularx}
\usepackage[utf8]{inputenc}
\usepackage{xcolor}  
\usepackage{colortbl} 
\usepackage{ragged2e} 
\usepackage{pifont}
\usepackage{enumitem}
\usepackage{cellspace}
\definecolor{tableHL}{RGB}{245,245,245} 

\DeclareMathOperator*{\argmin}{arg\, min}
\DeclareMathOperator{\Concat}{Concat}
\DeclareMathOperator{\PReLU}{PReLU}
\DeclareMathOperator{\Conv}{Conv}
\DeclareMathOperator{\GAP}{GAP}
\DeclareMathOperator{\Softmax}{Softmax}
\DeclareMathOperator{\DWConv}{DWConv} 
\DeclareMathOperator{\GELU}{GELU}

\DeclareUnicodeCharacter{2A02}{\otimes}

\newcommand{\moduleblock}[2]{\vspace{0.8em}\noindent\textbf{#1) #2}}

\newcolumntype{C}{>{\centering\arraybackslash}X}

\begin{document}

\title{Attention-Infused Autoencoder for Massive MIMO CSI Compression}

\author{Kangzhi Lou, \textit{Student Member, IEEE}, Xiping Wu, \textit{Senior Member, IEEE}

\thanks{The work was supported by the National Natural Science Foundation of China (NSFC). Kangzhi Lou acknowledges the support from the Beijing-Dublin International College Scholarship. This paper was published in part at 2025 IEEE Wireless Communications and Networking Conference (WCNC)\cite{Lou2025WCNC}. {\textit{Corresponding author: Xiping Wu.}}}

\thanks{K. Lou is with the School of Electrical and Electronic Engineering, University College Dublin, Belfield, Dublin, D04 V1W8, Republic of Ireland (e-mail: kangzhi.lou@ucdconnect.ie).}

\thanks{X. Wu is with the National Mobile Communications Research Laboratory, School of Information Science and Engineering, Southeast University, Nanjing 211189, China, and also with Purple Mountain Laboratories, Nanjing 211111, P.R.China (e-mail: xiping.wu@seu.edu.cn).}

}
\markboth{Journal of \LaTeX\ Class Files,~Vol.~XX, No.~XX, April~2025}%
{How to Use the IEEEtran \LaTeX \ Templates}

\maketitle
\begin{abstract}
As the number of multiple-input multiple-output (MIMO) antennas increases drastically with the development towards 6G systems, channel state information (CSI) compression becomes crucial for mitigating feedback overhead.
In recent years, learning models such as autoencoders (AE) have been studied for CSI compression, aiming to eliminate model assumptions and reduce compression loss. 
However, current learning methods are often designed and trained mainly for individual channel scenarios, with limited generalizability across different scenarios, of which the channel characteristics are prominently discrepant. 
Motivated by this, we propose a novel AE-based learning method named attention-infused autoencoder network (AiANet), which can parallelly and adaptively extract channel-wise and spatial features of CSI with an attention fusion mechanism. In addition, a locally-aware self-attention mechanism is developed to extract both global and local spatial patterns, to better capture the unique CSI features of different scenarios. Moreover, a mixed-training scheme is introduced to enable the proposed AiANet to gain generalizability across indoor and outdoor scenarios. 
Results show that when trained and tested in the same scenario, AiANet can substantially outperform the existing AE-based methods such as ACRNet, with an improvement of up to 3.42 dB in terms of normalized mean squared error (NMSE). With the mixed-training scheme, AiANet exhibits superior cross-scenario generalizability compared to the benchmark methods which are trained in one scenario and misused in another.
\end{abstract}


\section{Introduction}\label{sect:introduction}
\lettrine[loversize=0.1, nindent=0em]{T}{H}{E} vision of 6G networks, characterized by multi-Tbps data rates, ultra-low latency, and ultra-dense connectivity, promises transformative capabilities in wireless communications. Central to achieving this vision is massive MIMO technology, which employs base stations equipped with extensive antenna arrays to realize spectral efficiency gains of 10 to 100 times beyond 5G through advanced spatial precoding and wideband beamforming \cite{bjornson2017massive}.
The effectiveness of these techniques, however, critically depends on the availability of accurate downlink channel state information (CSI) at the base station. In Frequency Division Duplex (FDD) systems, acquiring this CSI necessitates feedback from the user.
As the number of antennas scales up, the volumn of CSI grows quadratically, leading to significant feedback overhead \cite{Hoydis2013Antenna}.
This overhead presents a major bottleneck, not only consuming precious uplink resources but also limiting the achievable precoding gain and thus capping the spectral efficiency benefits promised by massive MIMO. 
Consequently, efficient CSI compression is essential not just to mitigate this overhead, but to enable the timely and accurate channel knowledge required to unlock the full spectral efficiency and reliability potential of massive MIMO.

Traditional CSI compression methods, primarily based on compressed sensing (CS)\cite{candes2006near}, leverage the inherent sparsity of CSI in the angular-delay domain to reduce data dimensionality. 
CS maps high-dimensional CSI into a lower-dimensional space via projection with a measurement matrix, relying on the signal's sparsity in a chosen basis. The compressed data is then transmitted to the receiver, where the original CSI is reconstructed with high accuracy through sparsity-driven optimization techniques, such as Basis Pursuit \cite{chen2001atomic} and Orthogonal Matching Pursuit \cite{pati1993orthogonal}.
Notable works based on this principle include the LASSO $\ell_1$-solver\cite{daubechies2004iterative}, Approximate Message Passing (AMP)\cite{donoho2006compressed}, TVAL3\cite{li2009user}, and BM3D-AMP\cite{metzler2016denoising}. 
Despite their theoretical effectiveness, these methods typically assume strict sparsity conditions that rarely hold true in practical scenarios, resulting in suboptimal performance and limited applicability across diverse channel conditions. Moreover, their reliance on iterative optimization processes further restricts their effectiveness for rapidly varying wireless environments.

Recent advancements in deep learning, particularly autoencoder-based architectures, have shown superior performance in CSI compression tasks. The pioneering work, CsiNet\cite{wen2018deep}, significantly outperformed traditional methods despite its relatively simple structure. Building on this foundation, more sophisticated neural network models have emerged. For example, CRNet\cite{lu2020multi} adopts a multi-resolution architecture paired with an innovative training strategy to boost CSI compression performance. Subsequently, the same researchers introduced ACRNet\cite{lu2022binarized}, an enhanced model that incorporates network aggregation and a parametric rectified linear unit (PReLU) for further improvement. Around the same time, DCRNet\cite{tang2022dilated} was proposed, utilizing dilated convolutions within its residual blocks to effectively enlarge the receptive field and enhance the extraction of spatial features.
In \cite{ji2023enhancing}, a jigsaw puzzle framework was introduced as an auxiliary task to guide the training process, successfully enhancing compression performance. In addition, the study in \cite{liu2019bidirectional} leverages uplink CSI as an additional input to assist downlink CSI compression by exploiting the correlation between downlink and uplink channels.
Table~\ref{tab:csi_methods} displays some of the proposed methods for CSI compression.

\renewcommand{\arraystretch}{1.6}
\newcommand{\cmark}{\ding{51}}
\newcommand{\xmark}{\ding{55}}

\begin{table*}[ht]
\centering
\caption{Comparison of Recent CSI Feedback Compression Methods}
\label{tab:csi_methods}
\begin{tabular}{|>{\centering\arraybackslash}m{0.8cm}| 
                >{\centering\arraybackslash}m{2.0cm}| 
                >{\centering\arraybackslash}m{2.5cm}| 
                >{\centering\arraybackslash}m{2.0cm}| 
                >{\centering\arraybackslash}m{2cm}| 
                >{\raggedright\arraybackslash}m{6 cm}|} 
\specialrule{1pt}{0pt}{0pt}
\textbf{Year} & \textbf{Type} & \textbf{Method} & \textbf{BS Antenna Number} & \textbf{Cross-Scenario Compatibility} & \textbf{Remark} \\
\specialrule{0.5pt}{0pt}{0pt}
2006 & \multirow{2}{*}{\makecell{Compressed\\Sensing (CS)}} & LASSO \cite{donoho2006compressed} & - & \xmark & Sparse CSI modeling; simple, limited accuracy. \\\cline{3-6}
2016 &  & BM3D-AMP \cite{metzler2016denoising} & - & \xmark & Denoising + CS; computationally intensive. \\\hline
2018 & \multirow{8}{*}{\makecell{Convolutional\\Autoencoder}} & CsiNet \cite{wen2018deep} & 32 & \xmark & First DL-based CSI compression. \\\cline{3-6}
2020 &  & CRNet \cite{lu2020multi} & 32 & \xmark & Multi-resolution residual learning. \\\cline{3-6}
2021 &  & DeepCMC \cite{mashhadi2020distributed} & 16, 32, 64, 128 & \xmark & Distributed training, multi-antenna scalability. \\\cline{3-6}
2022 &  & ACRNet \cite{lu2022binarized} & 32 & \xmark & Aggregated convolution; binarization. \\\cline{3-6}
2023 &  & JigsawNet \cite{ji2023enhancing} & 32 & \xmark & Self-supervised training enhancement. \\\cline{3-6}
2024 &  & FSAMNet \cite{zhang2025fusion} & 32 & \xmark & Fusion self-attention for improved accuracy. \\\cline{3-6}
2024 &  & SwinCFNet \cite{cheng2024swin} & 32 & \xmark & Swin Transformer-based autoencoder. \\\cline{3-6}
2025 &  & AiANet (This Work) & 32 & \cmark & Mixed-training strategy to enhance generalization. \\
\specialrule{1pt}{0pt}{0pt}
\end{tabular}
\end{table*}

However, existing methods are specifically optimized for particular channel scenarios, such as indoor or outdoor environments. Consequently, a model trained for one scenario typically exhibits poor performance when applied to another, primarily due to substantial variations in CSI patterns arising from differing channel conditions. As a result, multiple scenario-specific models become necessary, significantly increasing both training complexity and computational overhead. Additionally, correctly identifying the channel scenario to select the appropriate model introduces further complexity and is prone to errors.
Nevertheless, developing a universal CSI compressor capable of effectively compressing CSI feedback from diverse environments poses significant challenges. The variations in CSI feedback arise from distinct multi-path propagation characteristics, mobility levels, spatial correlations, and interference patterns. Urban settings exhibit complex multi-path propagation and pronounced frequency selectivity, whereas rural or line-of-sight (LOS)-dominant scenarios typically feature simpler CSI structures. High-mobility conditions result in rapidly varying channels, whereas static environments maintain stable CSI with extended coherence times. 
These differing attributes complicate the development of a unified deep learning model. 
A robust CSI compressor should generalize effectively across diverse channel scenarios by capturing both common channel characteristics and fine-grained details.

To address these challenges, we introduce the Attention-infused Autoencoder Network (AiANet), a novel CSI compression model specifically designed for adaptive feature routing and extraction using attention-guided information flow to effectively capture both channel-wise and spatial correlations, thus improving reconstruction robustness under varied channel conditions.
In addition, a mixed-training strategy is proposed to promote model generalizability across scenarios.
Comprehensive simulation results demonstrate that AiANet consistently outperforms existing state-of-the-art autoencoder architectures for CSI compression with both separate-training and mixed-training, highlighting its practical applicability for real-world 6G deployments.
The main contributions of this work are as follows: 
\begin{itemize}
    \item Novel attention mechanisms (adaptive attention fusion and locally-aware self-attention) are proposed to effectively extract comprehensive channel-wise and spatial CSI features from both global and local contexts.
    \item A gated dense connection structure is introduced for effective multi-scale feature fusion and information flow control within the compression network.
    \item A robust training and evaluation framework is designed to assess and enhance model generalizability across wireless channel scenarios, including separate- and mixed-training, as well as intra- and cross-scenario testing.
\end{itemize}

The remainder of this paper is organized as follows. Section~\ref{sect:system_model} introduces the system model and underlying mathematical foundations. 
Section~\ref{sect:aianet} describes the proposed network architecture and the training and evaluation protocols. 
Ablation studies are presented in Section~\ref{sect:ablation_study}, followed by the simulation setup and performance evaluation in Section~\ref{sect:results}. 
Finally, Section~\ref{sect:conclusion} concludes the paper.

\section{System Model}
\label{sect:system_model}

\begin{figure*}[htbp] 
\centering
\includegraphics[width=\linewidth]{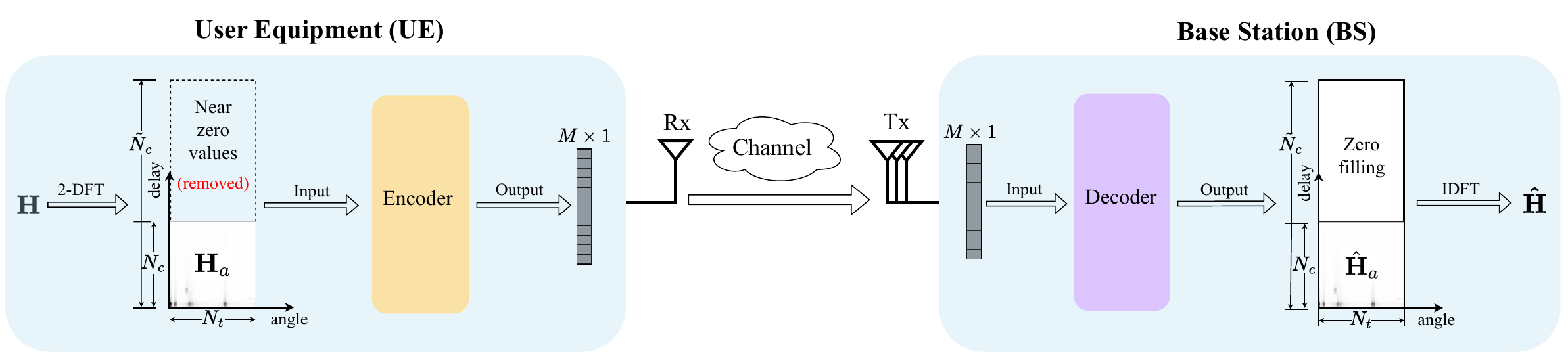}
\caption{Schematic diagram of the autoencoder-based CSI compression workflow.} 
\label{fig:CSI_transmission_workflow}
\end{figure*}

We consider a single-cell massive multiple-input multiple-output (MIMO) system operating in frequency division duplex (FDD) mode. A base station (BS) equipped with $N_t$ antennas ($N_t \gg 1$) serves a single-antenna user equipment (UE). The system utilizes $\Tilde{N_c}$ subcarriers for communication. 
On the $n$-th subcarrier ($n \in \{1, 2, \dots, \Tilde{N_c}\}$), the signal received by the UE can be modeled as:
 \begin{equation}
     y_n = {\mathbf{\Tilde{h}}}_n^H \mathbf{p}_n x_n + z_n,
 \end{equation}
where ${\mathbf{\Tilde{h}}}_n \in \mathbb{C}^{N_t \times 1}$ is the downlink channel vector for subcarrier $n$, $\mathbf{p}_n \in \mathbb{C}^{N_t \times 1}$ is the corresponding precoding vector applied at the BS, $x_n \in \mathbb{C}$ is the transmitted data symbol, and $z_n \in \mathbb{C}$ represents complex additive white Gaussian noise (AWGN) with zero mean and variance $\sigma_n^2$ (i.e., $z_n \sim \mathcal{CN}(0, \sigma_n^2)$).
The complete downlink CSI across all subcarriers forms the frequency-spatial CSI matrix:
 \begin{equation}
     {\mathbf{\Tilde{H}}} = [{\mathbf{\Tilde{h}}}_1, {\mathbf{\Tilde{h}}}_2, \dots , {\mathbf{\Tilde{h}}}_{\Tilde{N_c}}]^H \in \mathbb{C}^{\Tilde{N_c} \times N_t}.
 \end{equation}
Accurate knowledge of $\mathbf{\Tilde{H}}$ at the BS is crucial for optimizing the precoding vectors $\mathbf{p}_n$. In FDD systems, the BS obtains this information via feedback from the UE. The UE first estimates $\mathbf{\Tilde{H}}$ using downlink pilot signals transmitted by the BS and then sends this estimated CSI back to the BS through a limited-capacity uplink feedback channel.
Transmitting the full CSI matrix $\mathbf{\Tilde{H}}$ incurs significant overhead. As proposed in \cite{wen2018deep}, applying a 2-dimensional Discrete Fourier Transform (2D-DFT) can transform the CSI from the frequency-spatial domain to the angular-delay domain, resulting in a sparser representation. This transformation is given by:
\begin{equation} \label{eq: 2DFT}
    \mathbf{H} = \mathbf{F}_d \mathbf{\Tilde{H}} \mathbf{F}_a^H,
\end{equation}
where $\mathbf{F}_d \in \mathbb{C}^{\Tilde{N_c}\times\Tilde{N_c}}$ and $\mathbf{F}_a \in \mathbb{C}^{N_t\times N_t}$ are normalized DFT matrices. The resulting matrix $\mathbf{H} \in \mathbb{C}^{\Tilde{N_c} \times N_t}$ often exhibits significant energy concentration in the first few rows, corresponding to dominant path delays. By discarding the last $\Tilde{N_c} - N_c$ rows containing near-zero values (where $N_c \ll \Tilde{N_c}$ is chosen based on the maximum expected delay spread), the feedback size can be reduced. This yields a truncated angular-delay CSI matrix $\mathbf{H}_a \in \mathbb{C}^{N_c \times N_t}$.
The truncated matrix $\mathbf{H}_a$ consists of $N_c \times N_t$ complex values, equivalent to $2 N_c N_t$ real-valued numbers. In massive MIMO systems, where $N_t$ and $N_c$ can still be large (e.g., $\ge 32$), transmitting even the truncated CSI $\mathbf{H}_a$ represents substantial feedback overhead. Therefore, further compression of $\mathbf{H}_a$ is highly desirable.

While traditional compression techniques often rely on the sparsity of $\mathbf{H}_a$, this matrix may not be sufficiently sparse in practical scenarios where $N_t$ is finite \cite{wen2014channel}. Deep learning (DL)-based autoencoders \cite{rumelhart1986learning}, in contrast, can effectively learn complex, non-linear mappings to compress and reconstruct $\mathbf{H}_a$ with high fidelity, even if it is relatively dense. Autoencoders learn data-driven compression schemes, offering flexibility in adjusting the compression level and achieving superior reconstruction quality compared to traditional methods relying purely on sparsity.
Fig.~\ref{fig:CSI_transmission_workflow} outlines the autoencoder-based pipeline. At the UE, the original frequency-spatial CSI $\mathbf{\Tilde{H}}$ is transformed via 2D-DFT (Eq.~\eqref{eq: 2DFT}), and truncated to obtain $\mathbf{H}_a$. This matrix $\mathbf{H}_a$ (typically after separating real and imaginary parts into two channels) is fed into the encoder network $f_\mathcal{E}$. The encoder compresses $\mathbf{H}_a$ into a low-dimensional codeword vector $\mathbf{s} \in \mathbb{R}^M$, where $M = \eta \times (2 N_c N_t)$ and $\eta$ is the compression ratio. This codeword $\mathbf{s}$ is quantized, modulated, and transmitted over the uplink channel.

At the BS, the received codeword (potentially corrupted by noise, though often treated as noise-free in initial designs) is fed into the decoder network $f_\mathcal{D}$. The decoder reconstructs an estimate of the truncated angular-delay CSI, denoted $\hat{\mathbf{H}}_a$. This matrix is then zero-padded back to the original dimension $\Tilde{N_c} \times N_t$. Finally, an inverse 2D-DFT (2D-IDFT) is applied to obtain the reconstructed frequency-spatial CSI matrix $\hat{\mathbf{H}}_a$.

The core autoencoder process (encoding and decoding of the angular-delay representation) can be expressed as:
\begin{equation}
    \hat{\mathbf{H}}_a = f_\mathcal{D}(f_\mathcal{E}(\mathbf{H}_a; \Theta_\mathcal{E}); \Theta_\mathcal{D}),
\end{equation}
where $f_\mathcal{E}$ and $f_\mathcal{D}$ represent the encoder and decoder functions, parameterized by their respective learnable parameter sets $\Theta_\mathcal{E}$ and $\Theta_\mathcal{D}$. The training objective is to find the optimal parameters $(\hat{\Theta}_\mathcal{E}, \hat{\Theta}_\mathcal{D})$ that minimize the reconstruction error between the original truncated CSI $\mathbf{H}_a$ and the reconstructed version $\hat{\mathbf{H}}_a$. This is typically formulated as minimizing the Mean Squared Error (MSE), equivalent to minimizing the squared Frobenius norm:
\begin{equation}
    (\hat{\Theta}_\mathcal{E}, \hat{\Theta}_\mathcal{D}) = \argmin_{\Theta_\mathcal{E}, \Theta_\mathcal{D}} \|\mathbf{H}_a - \hat{\mathbf{H}}_a\|_F^2,
\end{equation}
where $\|\cdot\|_F$ denotes the Frobenius norm.

\section{Proposed Method}\label{sect:aianet}
This section presents the proposed AiANet, detailing its encoder and decoder architectures, along with training and evaluation methodologies. 
The system addresses CSI feedback compression through three key innovations: 1) Hybrid Attention-Gated Fusion (HAGF) for dynamic feature integration, 2) Locally-Aware Self-Attention (LASA) for multi-scale contextual modeling, and 3) Gated Dense Connections (GDC) for gradient flow optimization and feature refinement.
The encoder specifically preserves angular-delay domain characteristics through these modules, while the decoder ensures accurate CSI reconstruction from compressed codewords.


\begin{figure*}[htbp]
\centering
\includegraphics[width=\linewidth]{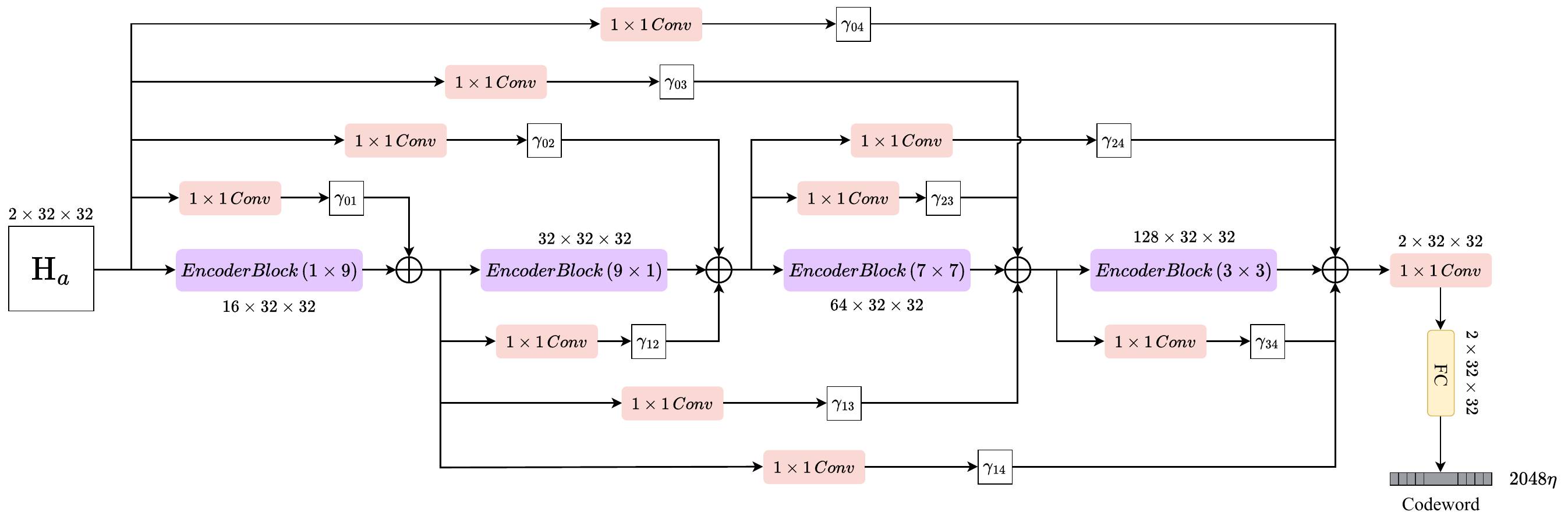}
\caption{Architecture of the AiANet encoder. The encoder processes the input through four sequential \textit{EncoderBlock}s with varied kernel sizes. As the data flows through these blocks, the number of channels increases (from 16 to 128), while the spatial resolution maintains the same ($32\times32$). 
Beyond the main processing path, the input to each EncoderBlock is also propagated --- scaled by learnable weights $\gamma_{ij}$ --- to its own output as well as the outputs of all subsequent EncoderBlocks.
The aggregated features from the final stage pass through a $1\times1~Conv$ and a Fully Connected (FC) layer to generate the codeword.}
\label{fig:aia_encoder}
\end{figure*}

\subsection{Encoder}
\begin{figure}[htbp]
\centering
\includegraphics[width=\linewidth]{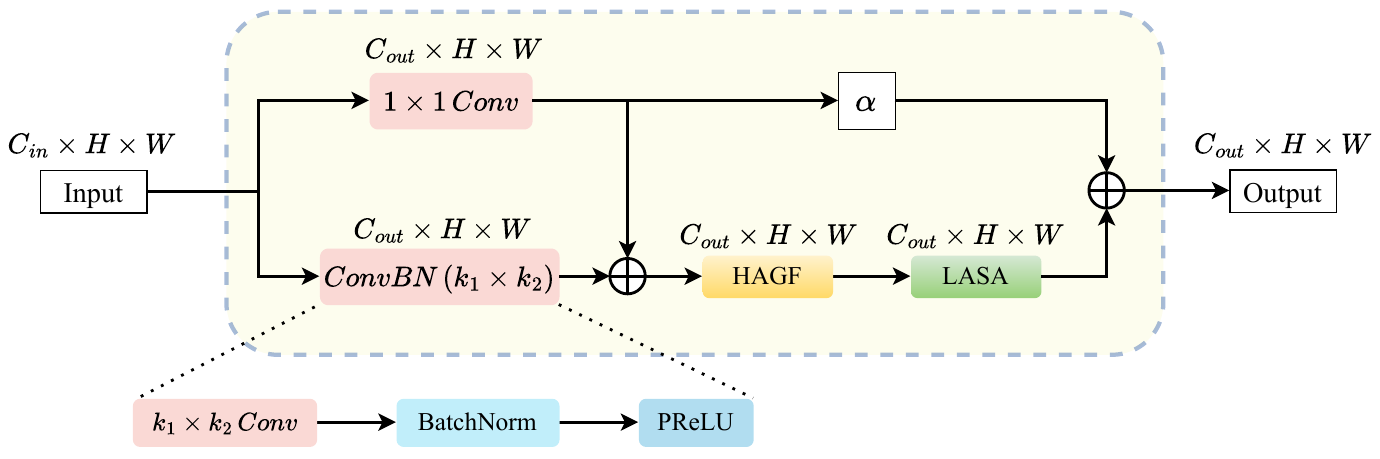}
\caption{The structure diagram of the EncoderBlock with a convolutional kernel of size $k_1 \times k_2$.  A learnable scalar parameter $\alpha$ is used to balance the contribution of the residual (upper) and main (lower) paths.}
\label{fig:encoderblock}
\end{figure}

The proposed encoder architecture, as shown in Fig.~\ref{fig:aia_encoder}, performs hierarchical CSI feature extraction through four cascaded EncoderBlocks (Fig.~\ref{fig:encoderblock}). 
Each EncoderBlock features two complementary processing paths: a main path employing multi-scale convolutions followed by attention refinement, and a residual path ensuring original feature preservation via adaptive projection. The convolutional kernels are specifically designed to target distinct angular-delay domain characteristics:
\begin{itemize}
\item $1\times9$ or $9\times1$ kernels: Capturing narrow horizontal or vertical strip features.
\item $7\times7$ kernels: Detecting distributions of large scattering clusters.
\item $3\times3$ kernels: Extracting fine-grained CSI details.
\end{itemize}
The final fully connected (FC) layer generates compressed codewords of dimension $2048\eta$, with $\eta$ denoting the compression ratio.

\moduleblock{1}{Hybrid Attention-Gated Fusion}

\begin{figure}[htbp]
\centering
\includegraphics[width=\linewidth]{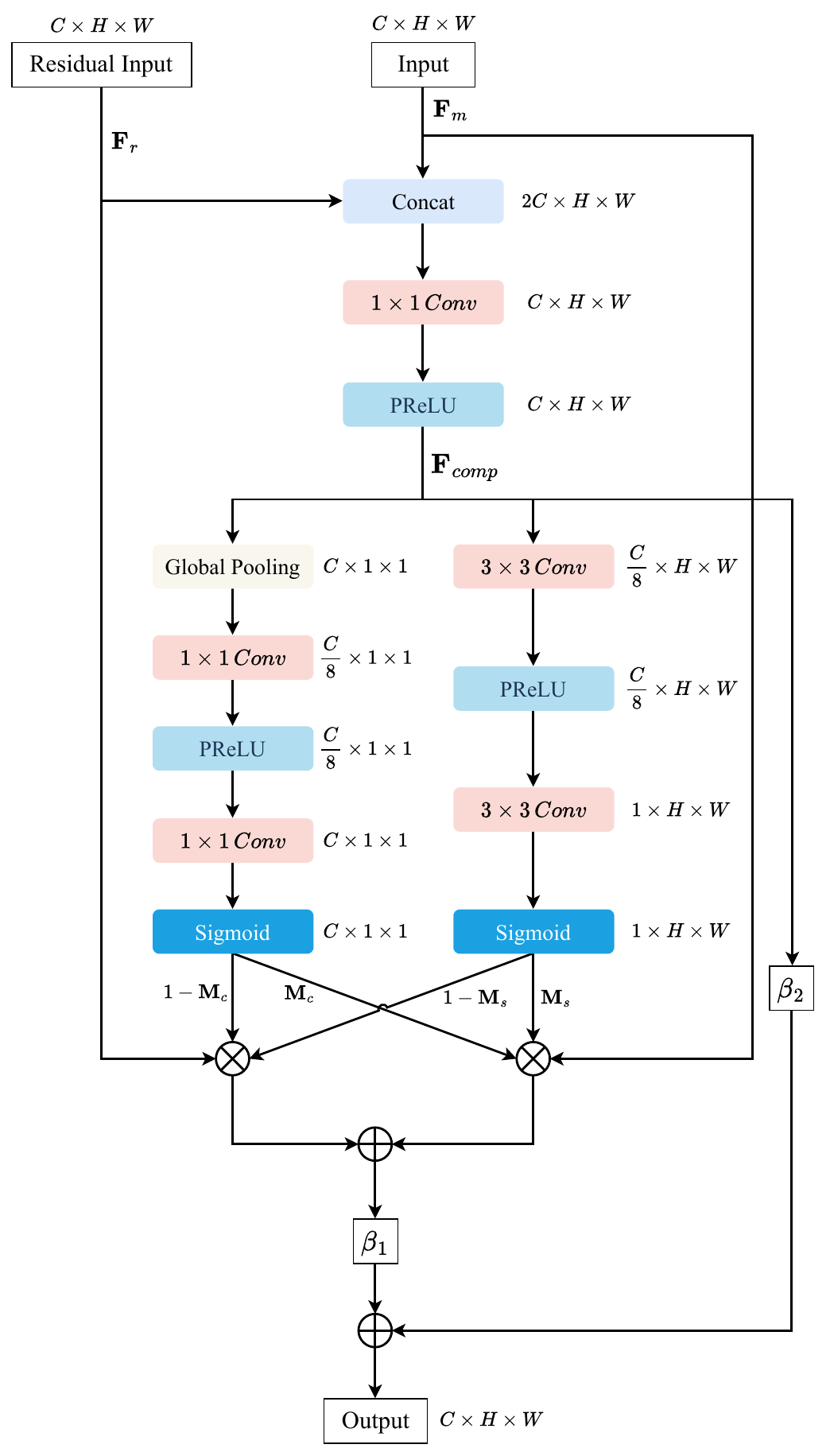}
\caption{Architecture of the Hierarchical Attention Gating Fusion (HAGF) module. It takes an Input feature map ($\mathbf{F}_m$) and a Residual Input ($\mathbf{F}_r$), processes them to generate complementary channel ($\mathbf{M}_c$) and spatial ($\mathbf{M}_s$) attention maps via parallel branches. A cross-gating mechanism applies these maps to refine features, which are then adaptively combined with the original Input ($\mathbf{F}_m$) using learnable parameters ($\beta_1$, $\beta_2$). }
\label{fig:hybrid_gate}
\end{figure}

The Hybrid Attention-Gated Fusion (HAGF) module (Fig.~\ref{fig:hybrid_gate}) dynamically integrates spatial and channel-wise features via dual parametric pathways. Through learnable gating mechanisms, it adaptively balances spatial and channel attention, prioritizing spatial attention in scattering-rich environments and emphasizing channel attention under frequency-selective fading conditions. The parallel arrangement of spatial and channel attention paths mitigates gradient conflicts, leading to smoother and more efficient model optimization.

HAGF initially concatenates residual ($\mathbf{F}_r$) and main ($\mathbf{F}_m$) features, forming a combined tensor $\mathbf{Z} \in \mathbb{R}^{2C \times H \times W}$. This tensor undergoes a $1\times1$ convolution and PReLU activation to produce compressed fusion features, $\mathbf{F}_{\text{comp}}$. Subsequently, channel-wise attention is computed via global average pooling (GAP), a dimensionality-reducing $1\times1$ convolution (reduction factor of 8), followed by PReLU activation, another expansion convolution, and sigmoid activation, generating the channel mask $\mathbf{M}_c$. Concurrently, spatial-wise attention involves sequential $3\times3$ convolutions that reduce the feature map to a single channel, also applying PReLU and sigmoid activation to yield the spatial mask $\mathbf{M}_s$. These masks act as soft selectors: 
the main features $\mathbf{F}_m$ are modulated by the joint mask $\mathbf{M} = \mathbf{M_c} \odot \mathbf{M_s}$ to emphasize co-activated channel-spatial patterns, while the residual features $\mathbf{F}_r$ are weighted by the complementary mask $\Bar{\mathbf{M}} = (\mathbf{1} - \mathbf{M}_c) \odot (\mathbf{1} - \mathbf{M}_s) \big)$ to preserve auxiliary information. 
In parallel, the compressed fusion feature $\mathbf{F}_{\text{comp}}$, derived before attention modulation, is incorporated via a residual pathways as a learnable scaled term. This auxiliary branch preserves mid-level representations of $\mathbf{F}_r$ and $\mathbf{F}_m$, thereby improving representational robustness and enabling smoother gradient propagation across layers.
The final fused output is thus computed as:
\begin{equation} \label{eq:hagf_fusion}
\mathbf{F}_{\text{fused}} = \beta_1 \big( \mathbf{F}_m \odot \mathbf{M}  + \mathbf{F}_r \odot \Bar{\mathbf{M}} \big) + \beta_2 \mathbf{F}_{\text{comp}},
\end{equation}
where $\beta_1$ and $\beta_2$ are learnable weights initialized with constraints. Alg.~\ref{alg:hagf_forward} provides further details on the HAGF operation.

\begin{algorithm}[htbp]
\small 
\caption{Forward Pass of the Hybrid Attention-Gated Fusion (HAGF)}
\label{alg:hagf_forward}

\KwIn{Residual features $\mathbf{F}_r \in \mathbb{R}^{C \times H \times W}$; main features $\mathbf{F}_m \in \mathbb{R}^{C \times H \times W}$}
\KwOut{Fused output features $\mathbf{F}_{\text{fused}} \in \mathbb{R}^{C \times H \times W}$}

\BlankLine 

\Begin{
    $\mathbf{Z} \leftarrow \Concat(\mathbf{F}_r, \mathbf{F}_m)$\; \tcp{$\mathbf{Z} \in \mathbb{R}^{2C \times H \times W}$}
    $\mathbf{F}_{\text{comp}} \leftarrow \PReLU(\Conv_{1\times1}(\mathbf{Z}))$\; \tcp{Feature compression}
    $\mathbf{M}_c \leftarrow \sigma\left(\Conv_{1\times1}\left(\PReLU\left(\Conv_{1\times1}(\GAP(\mathbf{F}_{\text{comp}}))\right)\right)\right)$\; \tcp{Channel attention mask}
    $\mathbf{M}_s \leftarrow \sigma\left(\Conv_{3\times3}\left(\PReLU\left(\Conv_{3\times3}(\mathbf{F}_{\text{comp}})\right)\right)\right)$\; \tcp{Spatial attention mask}
    $\hat{\mathbf{F}}_m \leftarrow \mathbf{F}_m \odot \mathbf{M}_c \odot \mathbf{M}_s$\; \tcp{Select attended main features}
    $\hat{\mathbf{F}}_r \leftarrow \mathbf{F}_r \odot (1 - \mathbf{M}_c) \odot (1 - \mathbf{M}_s)$\; \tcp{Select complementary residual features}
    $\mathbf{F}_{\text{fused}} \leftarrow \beta_1 (\hat{\mathbf{F}}_m + \hat{\mathbf{F}}_r) + \beta_2 \mathbf{F}_{\text{comp}}$\; \tcp{Features fusion}
    \KwRet $\mathbf{F}_{\text{fused}}$\; 
} 
\end{algorithm}

In Section~\ref{subsect: ablation_study_HAGF}, our ablation study confirms that HAGF's parallel architecture effectively mitigates semantic bias compared to sequential fusion methods like CBAM~\cite{woo2018cbam}, thus improving adaptive CSI reconstruction.

\moduleblock{2}{Locally-Aware Self-Attention}

\begin{figure*}[ht!]
\centering
\includegraphics[width=\textwidth]{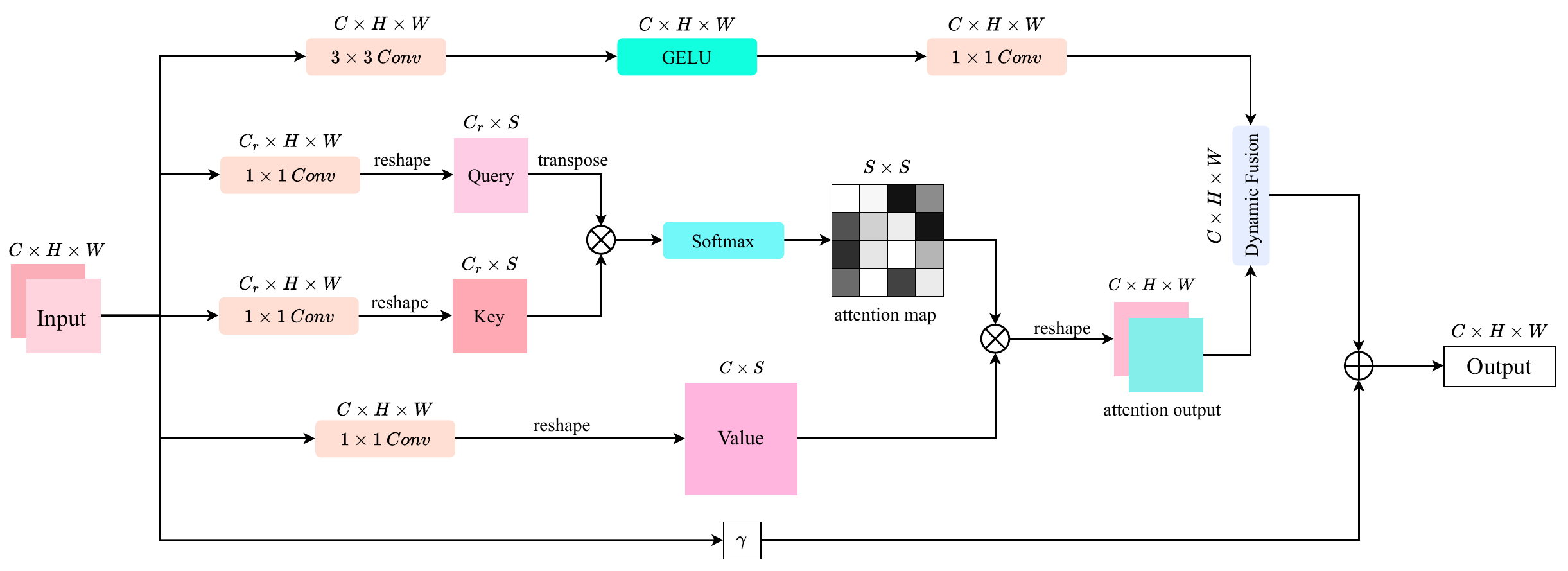}
\caption{Architecture of the Locally-Aware Self-Attention (LASA) module. It integrates features via three paths:(1) An upper path captures local patterns using convolutional layers ($3\times3$ Conv, $1\times$1 Conv). (2) A central path computes global self-attention using Query, Key, and Value projections. (3) A residual path provides a direct input connection scaled by $\gamma$. Outputs from the local and attention paths are merged by a Dynamic Fusion block, and the result is added to the scaled residual input (Note: In the attention path, $C_r$ denotes the reduced channel dimension; S = H * W).}
\label{fig:lasa}
\end{figure*}

To effectively balance global and local spatial correlations, we propose the Locally-Aware Self-Attention (LASA) module, which combines local feature preservation with global context modeling capabilities inherent in self-attention mechanisms. As illustrated in Fig.~\ref{fig:lasa}, LASA adopts a multi-branch architecture composed of a global self-attention module (middle three branches), a local context enhancement path (upper branch), and a residual connection integrated with an adaptive fusion gate (lower branch).

The global self-attention branch begins by applying parallel $1\times1$ convolutions to the input feature map $\mathbf{X}{in} \in \mathbb{R}^{C \times H \times W}$ to generate the query, key, and value tensors. Specifically, the query ($\mathbf{Q}$) and key ($\mathbf{K}$) tensors are projected to a lower-dimensional space $C_r = C/4$ to reduce computational cost, while the value tensor ($\mathbf{V}$) retains full channel resolution. The global attention is computed via scaled dot-product attention:
\begin{equation}
    \mathbf{F}_{\text{att}} = \text{Reshape}\left(\text{Softmax}\left(\frac{\mathbf{Q}^\top\mathbf{K}}{\sqrt{C_r}}\right)\mathbf{V}^\top\right),
\end{equation}
where $\mathbf{F}_{\text{att}} \in \mathbb{R}^{C\times H\times W}$.
In parallel, the local enhancement path applies a $3\times3$ depthwise convolution followed by GELU activation and a $1\times1$ pointwise convolution to yield locally refined features:
\begin{equation}
\mathbf{F}_{\text{local}} = \text{Conv}_{1\times1}\left(\text{GELU}\left(\text{Conv}_{3\times3}(\mathbf{X}{in})\right)\right).
\end{equation}
This path is designed to retain crucial spatial patterns such as delay-bin correlations, near-field phase coherence, and orientation sensitivity.
The two feature streams are then fused by a learnable channel-wise gate through a sigmoid-based soft selection mechanism:
\begin{equation}
\mathbf{F}_{\text{fused}} = \sigma(\omega) \odot \mathbf{F}_{\text{att}} + (\mathbf{1} - \sigma(\omega)) \odot \mathbf{F}_{\text{local}},
\end{equation}
where $\omega \in \mathbb{R}^C$ is a learnable parameter, $\sigma(\cdot)$ is the sigmoid function, and $\odot$ denotes element-wise multiplication. This gating process allows the network to adaptively balance the contributions of local and global information.
Finally, a residual connection scaled by a trainable scalar $\gamma$ (initialized to 0.1) is added to preserve original input information and facilitate gradient propagation:
\begin{equation}
\mathbf{F}_{\text{out}} = \mathbf{F}_{\text{fused}} + \gamma \mathbf{X}_{\text{in}}.
\end{equation}

Alg.~\ref{alg:lasa} outlines the forward procedure of the LASA module. Overall, LASA enhances wireless signal representation by capturing both fine-grained spatial patterns and long-range global dependencies with low overhead and high adaptability.

\begin{algorithm}[htbp]
\small 
\caption{Forward Pass of the Locally-Aware Self-Attention (LASA)}
\label{alg:lasa}

\KwIn{Input feature map $\mathbf{X}_{\text{in}} \in \mathbb{R}^{C \times H \times W}$}
\KwOut{Output feature map $\mathbf{F}_{\text{out}} \in \mathbb{R}^{C \times H \times W}$}
\BlankLine 

\Begin{
    $\mathbf{Q} \leftarrow \Conv_{1\times1}^{C \to C/4}(\mathbf{X}_{\text{in}})$\; \tcp{Query projection}
    $\mathbf{K} \leftarrow \Conv_{1\times1}^{C \to C/4}(\mathbf{X}_{\text{in}})$\; \tcp{Key projection}
    $\mathbf{V} \leftarrow \Conv_{1\times1}^{C \to C}(\mathbf{X}_{\text{in}})$\; \tcp{Value projection}

    $\mathbf{A} \leftarrow \Softmax\left(\frac{\mathbf{Q}\mathbf{K}^\top}{\sqrt{C/4}}\right)$\; \tcp{Scaled attention weights}
    $\mathbf{F}_{\text{att}} \leftarrow \mathbf{A}\mathbf{V}^\top$\; \tcp{Global context aggregation}

    $\mathbf{F}_{\text{dw}} \leftarrow \DWConv_{3\times3}(\mathbf{X}_{\text{in}})$\; \tcp{Depthwise convolution}
    $\mathbf{F}_{\text{local}} \leftarrow \GELU(\Conv_{1\times1}(\mathbf{F}_{\text{dw}}))$\; \tcp{Local context}

    $\mathbf{F}_{\text{fused}} \leftarrow \sigma(\omega)\odot\mathbf{F}_{\text{att}} + (1-\sigma(\omega))\odot\mathbf{F}_{\text{local}}$\; \tcp{Fusion}
    $\mathbf{F}_{\text{out}} \leftarrow \mathbf{F}_{\text{fused}} + \gamma \mathbf{X}_{\text{in}}$\; \tcp{Residual connection}

    \KwRet $\mathbf{F}_{\text{out}}$\; 
} 
\end{algorithm}

\moduleblock{3}{Gated Dense Connection}

To facilitate multi-scale feature fusion and optimize gradient flow, we improved the  DenseNet\cite{zhu2017densenet} by embedding a learnable scaling factor in each residual connection, which we call the gated dense connection (GDC).
As illustrated in Fig.~\ref{fig:aia_encoder}, the encoder comprises four sequentially cascaded EncoderBlocks, which progressively expand feature dimensions from 2 to 128 channels (2 $\rightarrow$ 16 $\rightarrow$ 32 $\rightarrow$ 64 $\rightarrow$ 128). These EncoderBlocks are interconnected through our novel adaptive gating mechanisms, promoting efficient multi-scale feature propagation.

Within the encoder architecture, each block integrates multi-level feature inputs through three distinct connection strategies. Firstly, direct hierarchical transmissions pass outputs from Block$_n$ directly into Block$_{n+1}$, preserving immediate spatial context. Secondly, cross-layer dense connections deliver transformed initial input features directly into subsequent blocks (Input $\rightarrow$ Block$_n$ for $n > 1$), enhancing long-range information exchange. Lastly, lateral feature reinforcements allow features from preceding non-adjacent blocks (Block$_m$ where $m < n-1$) to supplement later stages (Block$_n$), enriching the diversity of feature representations.
Formally, given an input tensor $\mathbf{X} \in \mathbf{R}^{2\times H\times W}$, the output of the $k$-th block, $O_k$, is computed as:
\begin{equation}
O_k = A_k\left(\sum_{i=1}^{k-1} W_i^{(k)} \cdot T_i^{(k)}(O_i) + W_x^{(k)} \cdot T_x^{(k)}(X)\right),
\end{equation}
where $A_k(\cdot)$ denotes the set of operations encapsulated in EncoderBlock$_k$. Channel-wise transformations are applied by $T_i^{(k)}(\cdot)$ using $1\times1$ convolutions to harmonize channel dimensions across blocks. The gating coefficients $W_i^{(k)} \in [0, 1]$ represent trainable parameters that adaptively scale the importance of connections, while $W_x^{(k)}$ serves as the projection weight from the input tensor.

The adaptive gating mechanism employs trainable parameters initialized to zero, subsequently modulated by sigmoid activation functions:
\begin{equation}
W = \sigma(\alpha) = \frac{1}{1 + e^{-\alpha}}.
\end{equation}
This design choice ensures stable initial training conditions (with initial gate values approximately 0.5), allowing the network to automatically select and emphasize meaningful pathways while simultaneously suppressing redundant feature flow. Such adaptive gating fosters dynamic recalibration of feature hierarchies throughout training.

In summary, the gated dense connectivity approach effectively integrates comprehensive features from all preceding hierarchical levels, thereby preserving spatial information integrity through carefully constrained projection operations. As validated by ablation studies in Section~\ref{subsect:ablation_study_HGDC}, this mechanism enhances network performance through two complementary effects: (1) dynamic emphasis on semantically significant feature streams, and (2) suppression of redundant information propagation. This dual functionality simultaneously improves training stability through gradient regularization while increasing representational efficiency.

\subsection{Decoder}
\label{subsect: AiANet_decoder}

\begin{figure}[htbp]
\centering
\includegraphics[width=0.8\linewidth]{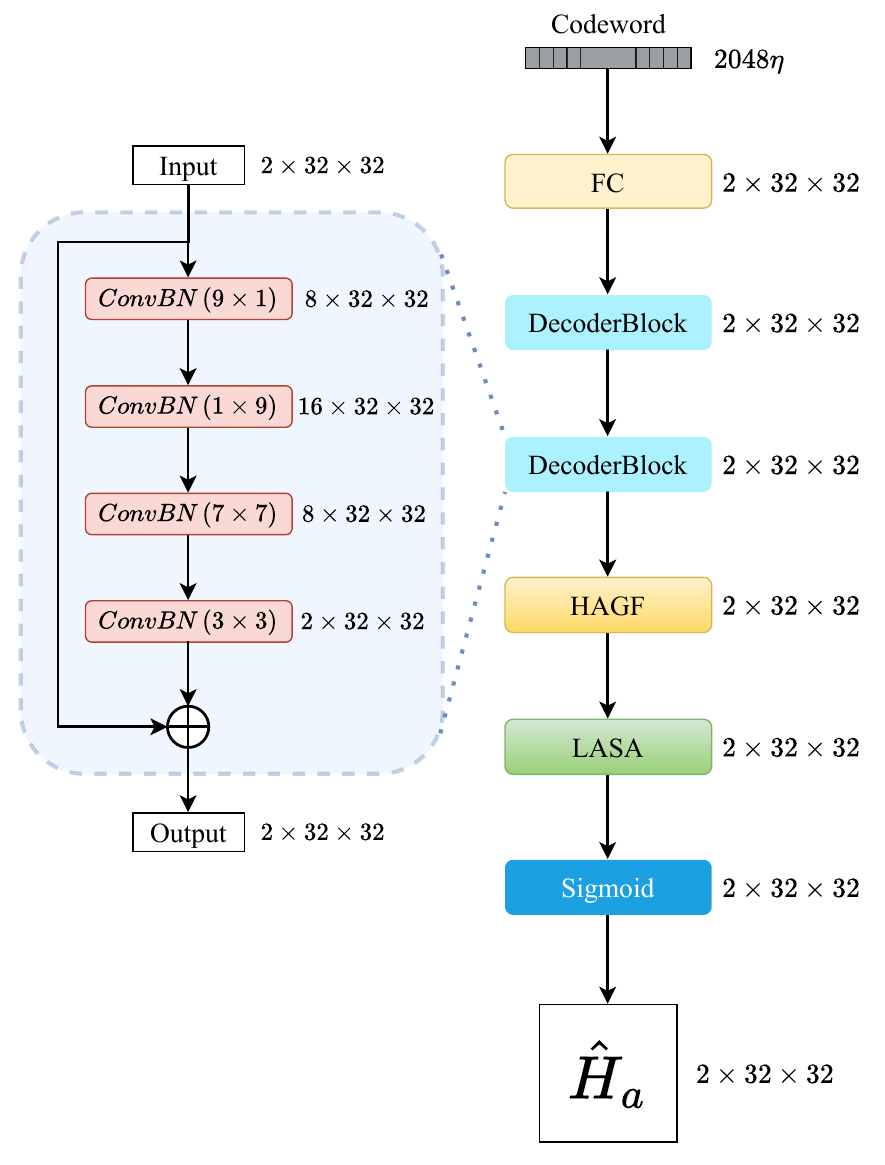}
\caption{The structure diagram of the decoder in AiANet.}
\label{fig:aia_decoder}
\end{figure}

The decoder of AiANet, as illustrated in Fig.~\ref{fig:aia_decoder}, employs a streamlined structure that strategically balances complexity with reconstruction capability. While architecturally simpler than the encoder, the decoder critically transforms compressed codewords into high-fidelity CSI reconstructions through three key processing stages.
The pipeline of AiANet decoder initiates with a compressed codeword of dimension 2048$\eta$ (where $\eta$ denotes the compression ratio), which is first projected to the original CSI dimensions ($2\times 32 \times 32$) via a fully connected layer. 
This base reconstruction undergoes progressive refinement through two cascaded DecoderBlocks --- each containing four cascaded ConvBN layers with kernel sizes {1, 3, 5, 7} for multi-scale feature processing.
Residual skip connections within each block mitigate gradient vanishing while enabling cross-level feature fusion.
Subsequent to the DecoderBlocks, an HAGF module and a LASA module cooperatively enhance feature discriminability through channel-wise recalibration and spatial dependency modeling.
The reconstruction pipeline culminates in a Sigmoid activation layer that normalizes outputs to (0, 1), producing the final estimated CSI matrix $\hat{\mathbf{H}}_a \in \mathbb{R}^{2 \times 32 \times 32}$ as shown in Fig.\ref{fig:aia_decoder}.

\subsection{Training and Evaluation Protocol}
\label{sect:protocol}
We adopt a commonly used CSI dataset generated by \cite{wen2018deep}, following the default COST2100 model settings \cite{liu2012cost}, to ensure fair comparisons.
The dataset contains indoor (5.3 GHz) and outdoor (300 MHz) scenarios, with CSI matrices sparsified via 2D-DFT in the angular-delay domain.
The base station employs $N_t = 32$ antennas and operates over multiple subcarriers. 
Due to multipath constraints, only the first $N_c = 32$ rows of the CSI matrix have non-zero entries; hence, these rows are retained to reduce the dimension to $2 \times N_c \times N_t$. 
Despite this reduction, the dataset size remains large for massive MIMO systems, necessitating further  compression. We split the data into training, validation, and test sets, comprising 100k, 30k, and 20k samples, respectively.

Reconstruction quality is quantified through two metrics: the Normalized Mean Square Error (NMSE) and Cosine Similarity ($\rho$).
These metrics represent the discrepancy between the original angular-delay domain CSI, $\mathbf{H}_a$, and its reconstructed counterpart, $\hat{\mathbf{H}}_a$.
The NMSE is defined as 
\begin{equation}
    \text{NMSE} = \mathbb{E} \left\{ \frac{\|\mathbf{H}_a - \hat{\mathbf{H}}_a\|_2^2}{\|\mathbf{H}_a\|_2^2} \right\},
    \label{eq:nmse}
\end{equation}
where the expectation is computed over the evaluation dataset. Lower NMSE indicates higher reconstruction fidelity. 
And the Cosine Similarity ($\rho$) is given by
\begin{equation}
    \rho = \mathbb{E} \left\{ \frac{\lvert\mathbf{H}_a^H \cdot \hat{\mathbf{H}}_a\rvert}{\|\mathbf{H}_a\|_2 \|\hat{\mathbf{H}}_a\|_2} \right\},
    \label{eq:cosine_similarity}
\end{equation}
where $\mathbf{H}_a^H$ represents the Hermitian (conjugate) transpose of $\mathbf{H}_a$. A higher $\rho$ reflects better alignment between $\mathbf{H}_a$ and $\hat{\mathbf{H}}_a$.

The simulation investigates performance across four compression ratios ($\eta$): {1/4, 1/16, 1/32, and 1/64}. Batch-level data shuffling is applied during training to ensure a robust and unbiased evaluation.
Two distinct training strategies are employed:

\begin{itemize}
\item \textbf{Separate Training}: 
Models are trained and validated using data exclusively from a single scenario (either indoor or outdoor). The evaluation for this strategy involves two testing methodologies:
    \begin{itemize}
            \item \textbf{Intra-scenario Testing:} Models are tested using data from the same scenario used during their training (e.g., an indoor-trained model tested on indoor data).
            \item \textbf{Cross-scenario Testing:} Models are tested using data from the other, previously unseen scenario (e.g., an indoor-trained model tested on outdoor data). This assesses the model's cross-scenario generalization capability.
    \end{itemize}
\item \textbf{Mixed Training}: 
Models are trained and validated on a dataset composed of an equal number (1:1 ratio) of indoor and outdoor samples. This approach aims to enhance inherent cross-scenario generalization. For evaluation, a balanced test set containing equal samples from both scenarios is used to comprehensively assess performance across diverse environments.
\end{itemize}

\section{Ablation Studies}
\label{sect:ablation_study}

This section systematically evaluates the contribution of key architectural components within AiANet via component-wise ablation.
All experiments were conducted under the mixed-training scheme, using the AdamW optimizer with a weight decay of \( 1 \times 10^{-4} \) and training for 1000 epochs.

\subsection{Effectiveness of HAGF}
\label{subsect: ablation_study_HAGF}

\begin{figure}[htbp]
\centering
\includegraphics[width=\linewidth]{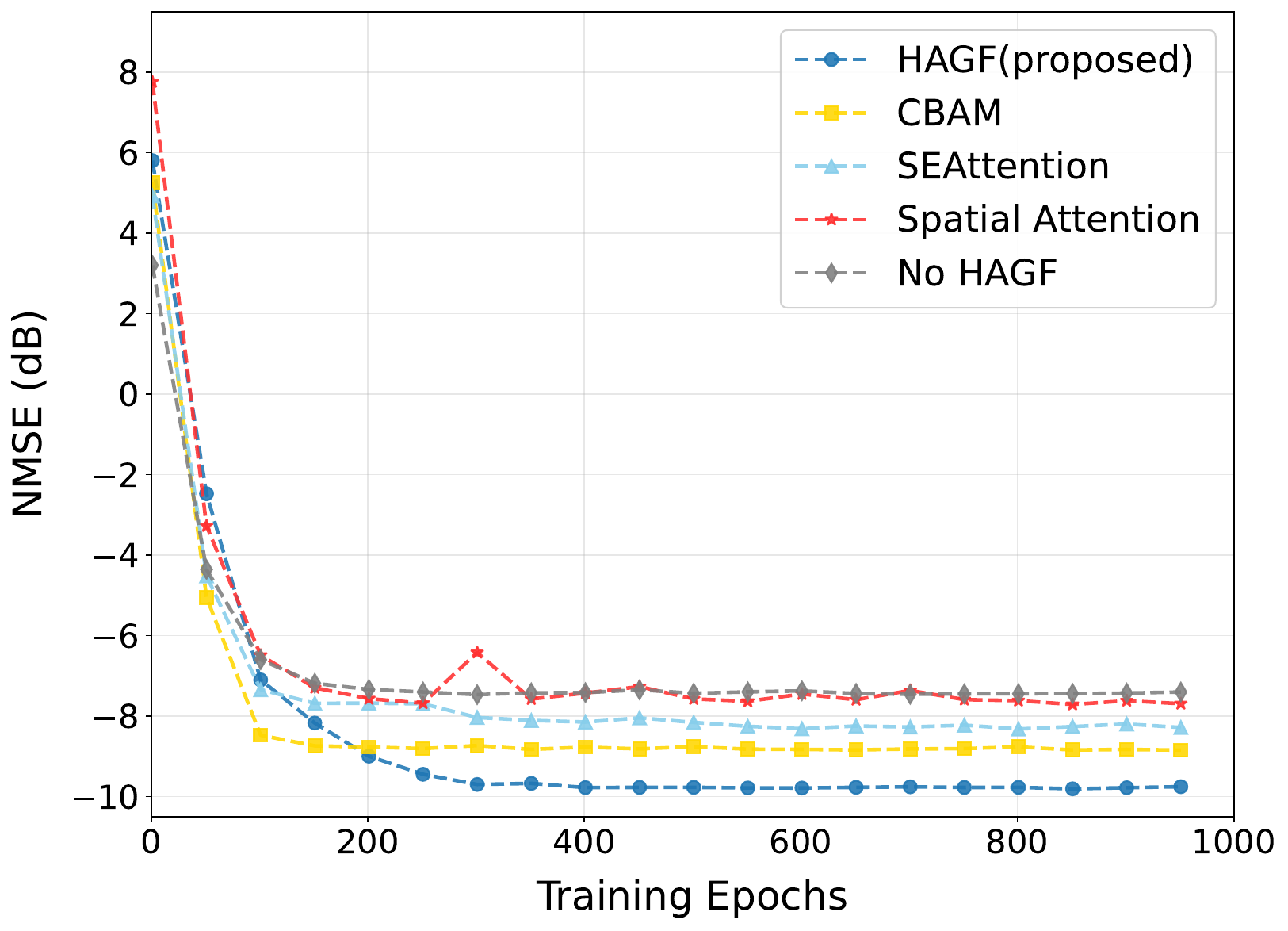}
\caption{Convergence comparison for attention module ablation ($\eta = 1/16$). NMSE (dB) versus training epochs for configurations: HAGF (Proposed), CBAM\cite{woo2018cbam}, SEAttention\cite{hu2018squeeze}, Spatial Attention\cite{jaderberg2015spatial}, and No Attention.}
\label{fig:ablation_study_HAGF}
\end{figure}

To assess the impact of the proposed HAGF module, we compare AiANet's performance when HAGF is included versus when it is replaced by alternative attention mechanisms or removed entirely.
The configurations evaluated are: 1) Proposed (HAGF), 2) CBAM \cite{woo2018cbam}, 3) SEAttention \cite{hu2018squeeze}, 4) Spatial Attention only (similar to the spatial branch in \cite{woo2018cbam, jaderberg2015spatial}), and 5) No-Attention baseline module.
Table~\ref{tab:hagf_ablation} shows the NMSE (in dB) performance of each configuration across compression ratios ($\eta = 1/4$ to $1/64$), and Fig.~\ref{fig:ablation_study_HAGF} tracks the convergence behavior over epochs under $\eta = 1/16$.

As shown in Table~\ref{tab:hagf_ablation}, the configuration with the proposed HAGF module consistently achieved the lowest NMSE values across all compression ratios. For instance, at $\eta = 1/4$, HAGF attains -15.24 dB, outperforming CBAM (-14.89 dB), SEAttention (-14.05 dB), Spatial Attention (-12.74 dB), and the No-Attention baseline (-11.92 dB). 
The performance advantage of HAGF over the No-Attention baseline decreases from 3.32 dB at $\eta = 1/4$ to 0.91 dB at $\eta = 1/64$. 
This diminishing gap is expected, as aggressive compression leads to significant information loss, limiting the potential gains from any attention mechanism.
Fig.~\ref{fig:ablation_study_HAGF} reveals HAGF's faster convergence, stabilizing at -9.8 dB by epoch 500 versus CBAM's -8.87 dB.
This demonstrates HAGF's dual-branch architecture effectively preserves channel-spatial feature integrity, unlike sequential attention mechanisms that introduce semantic bias.

\begin{table}[htbp]
    \centering
    \footnotesize
    \setlength{\tabcolsep}{3.2pt} 
    \renewcommand{\arraystretch}{1.3} 
    \caption{Performance Comparison of Hybrid Attention Mechanisms (NMSE in dB)}
    \label{tab:hagf_ablation}
    \begin{threeparttable}
        \begin{tabularx}{\columnwidth}{
            |>{\centering\arraybackslash}m{0.8cm}| 
            *{4}{>{\centering\arraybackslash}X|}  
            >{\centering\arraybackslash}X|         
            }
            \Xhline{2\arrayrulewidth} 
            \multirow{2}{*}[-6pt]{$\eta$} & \multicolumn{5}{c|}{Ablation Study -- HAGF} \\ 
            \cline{2-6}   
            & \makecell{HAGF} 
            & \makecell{CBAM\\\cite{woo2018cbam}} 
            & \makecell{\scriptsize SEAttention\\\cite{hu2018squeeze}} 
            & \makecell{\scriptsize Spatial\\ \scriptsize Attention\\\cite{jaderberg2015spatial}} 
            & \makecell{\scriptsize No\\ \scriptsize HAGF} \\ 
            \hline
            1/4  & \textbf{-15.24} & -14.89 & -14.05 & -12.74 & -11.92 \\ 
            1/16 & \textbf{-9.81}  & -8.87  & -8.40  & -7.73  & -7.12  \\ 
            1/32 & \textbf{-6.88}  & -6.53  & -6.47  & -6.12  & -5.84  \\ 
            1/64 & \textbf{-4.70}  & -4.46  & -4.36  & -4.03  & -3.79  \\ 
            \Xhline{2\arrayrulewidth} 
        \end{tabularx}
    \end{threeparttable}
\end{table}

\subsection{Effectiveness of LASA}
\label{subsect: ablation_study_self_attention}

\begin{figure}[ht]
\centering
\includegraphics[width=\linewidth]{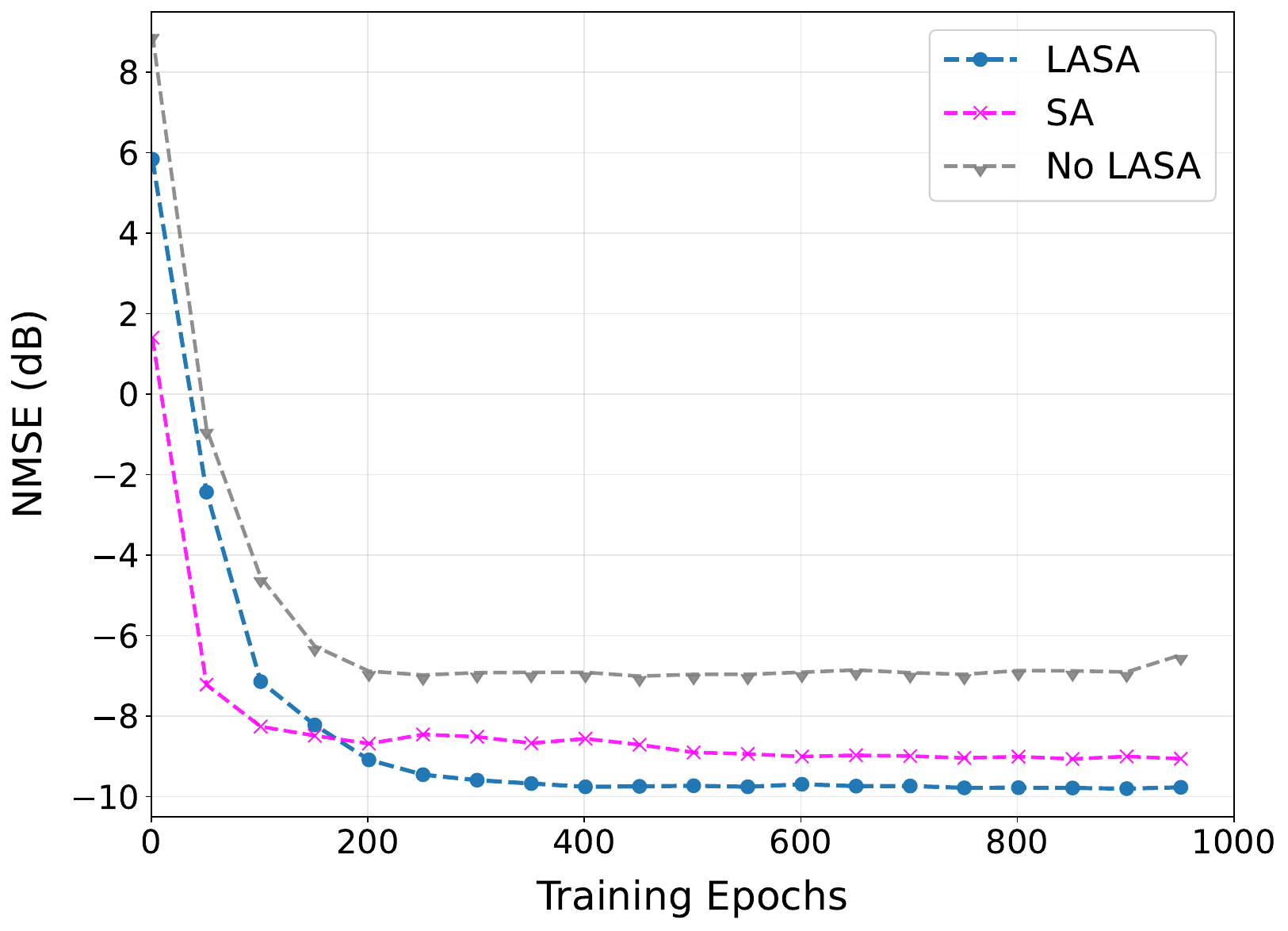} 
\caption{Convergence behavior (NMSE in dB vs. training epochs) for the LASA ablation study at $\eta=1/16$. Compared configurations are: the full model with LASA (`LASA'), a variant replacing LASA with standard Self-Attention (`SA'), and the baseline model with the module removed (`Baseline (Removed)').} 
\label{fig:ablation_study_lasa}
\end{figure}

To evaluate the effectiveness of the proposed LASA module, we conducted an ablation study comparing three distinct configurations: 1) the full model incorporating LASA, 2) a variant replacing LASA with standard Self-Attention (SA), and 3) a baseline model with the module entirely removed.  
Table~\ref{tab:ablation_study_lasa} presents the NMSE results of the above three configurations across four compression ratios, while Fig.~\ref{fig:ablation_study_lasa} illustrates the convergence behavior over 1,000 training epochs at $\eta = 1/16$.

Table~\ref{tab:ablation_study_lasa} clearly shows that the LASA configuration achieved the lowest NMSE (i.e., best performance) across all tested compression ratios. For instance, at the most aggressive compression ($\eta = 1/64$), the LASA model achieved -4.70 dB, outperforming the SA variant (-4.22 dB) and the baseline (-3.93 dB). Similar to trends observed in HAGF, the performance advantage of LASA over the baseline diminished as compression increased. The NMSE gap decreased from 2.97 dB at $\eta = 1/4$ (LASA: -15.24 dB vs. Baseline: -12.27 dB) to 0.77 dB at $\eta = 1/64$. This suggests that at very high compression levels, the inherent difficulty of the reconstruction task limits the achievable gains from any specific architecture enhancement.

The convergence analysis in Fig.~\ref{fig:ablation_study_lasa} (for $\eta = 1/16$) provides further insight. The LASA configuration demonstrated stable convergence, reaching approximately -9.81 dB by epoch 300. In contrast, the baseline model plateaued much earlier at a significantly higher NMSE of about -7.33 dB. Although the SA variant exhibited faster initial convergence (before epoch 80), it suffered from subsequent performance fluctuations and ultimately converged to -9.07 dB, roughly 0.74 dB worse than the LASA model.
Collectively, these results demonstrate that the specific design of the LASA module plays a crucial role in enhancing CSI compression and reconstruction performance compared to both standard self-attention and a simpler baseline architecture.

\begin{table}[htbp]
    \centering
    \footnotesize 
    \setlength{\tabcolsep}{3.2pt} 
    \caption{NMSE (dB) Performance Comparison for LASA Ablation Study at Different Compression Ratios ($\eta$).} 
    \label{tab:ablation_study_lasa}
    \renewcommand{\arraystretch}{1.3} 
    \begin{threeparttable}
        \begin{tabularx}{\columnwidth}{
        |>{\centering\arraybackslash}m{0.8cm}|>{\centering\arraybackslash}X|>{\centering\arraybackslash}X|>{\centering\arraybackslash}X|}
        \Xhline{2\arrayrulewidth}
        \multirow{2}{*}{\vspace{0.1cm} $\eta$} & \multicolumn{3}{c|}{Ablation Study -- LASA} \\ \cline{2-4} 
         & LASA & Self-Attention (SA) & No LASA \\ \hline 
        1/4  & \textbf{-15.24} & -13.92 & -12.27 \\
        1/16 & \textbf{-9.81}  & -9.07  & -7.33  \\
        1/32 & \textbf{-6.88}  & -6.07  & -5.58  \\
        1/64 & \textbf{-4.70}  & -4.22  & -3.93  \\
        \Xhline{2\arrayrulewidth}
        \end{tabularx}
    \end{threeparttable}
\end{table}

\subsection{Effectiveness of the Gated Dense Connections} 
\label{subsect:ablation_study_HGDC}

The AiANet encoder employs Gated Dense Connections (GDC), a specialized skip connection mechanism designed to enhance feature reuse and mitigate potential vanishing gradients. Unlike traditional DenseNet connections, GDC introduces learnable scaling factors (gates) to dynamically adjust the contribution of skip connections and utilizes additive merging instead of concatenation to maintain dimensionality.

\begin{figure}[htbp]
\centering
\includegraphics[width=\linewidth]{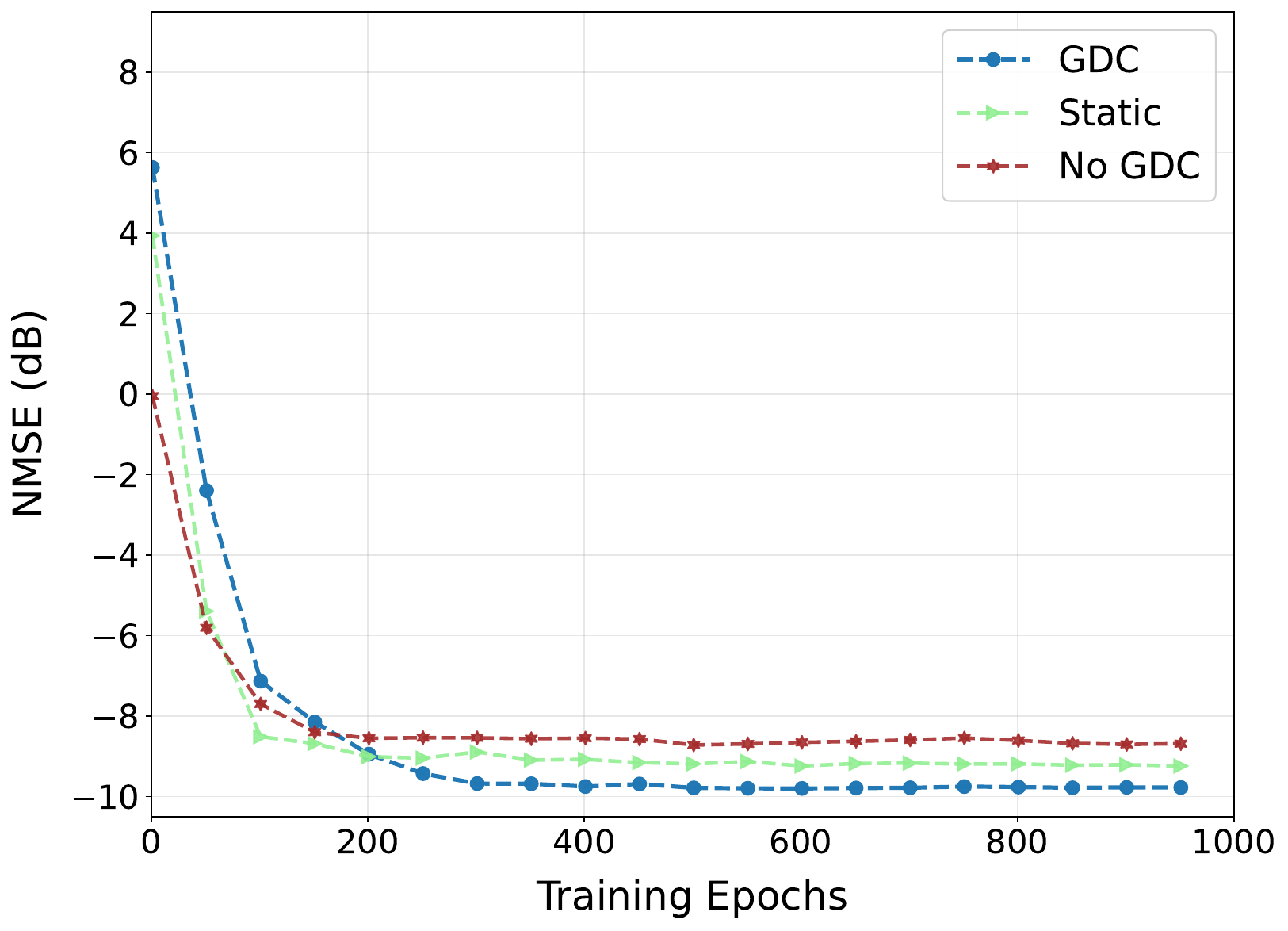} 
\caption{Convergence behavior (NMSE in dB vs. training epochs) for the GDC ablation study at $\eta=1/16$. Compared configurations are: the full model with GDC (`GDC'), a variant with static dense connections (`Static'), and the baseline model without dense connections (`Baseline (Removed)').} 
\label{fig:ablation_study_SDC}
\end{figure}

To isolate and evaluate the contribution of GDC, we performed an ablation study comparing three encoder configurations: 1) the proposed architecture using GDC, 2) a variant employing static dense connections without gating (Static), and 3) a baseline model lacking any dense skip connections (Baseline (Removed)). We assessed performance using NMSE (dB) across compression ratios ($\eta$) from 1/4 to 1/64, reported in Table~\ref{tab:encoder_SDC}, and analyzed convergence behavior over 1,000 training epochs at $\eta = 1/16$, shown in Fig.~\ref{fig:ablation_study_SDC}.

\begin{table}[htbp]
    \centering
    \footnotesize
    \setlength{\tabcolsep}{3.2pt}
    \caption{NMSE (dB) Performance Comparison for GDC Ablation Study at Different Compression Ratios ($\eta$).} 
    \label{tab:encoder_SDC}
    \renewcommand{\arraystretch}{1.3}
    \begin{threeparttable}
        \begin{tabularx}{\columnwidth}{|>{\centering\arraybackslash}m{0.8cm}|>{\centering\arraybackslash}X|>{\centering\arraybackslash}X|>{\centering\arraybackslash}X|}
        \Xhline{2\arrayrulewidth}
        \multirow{2}{*}{\vspace{0.1cm}\(\eta\)} & \multicolumn{3}{c|}{Ablation Study -- GDC} \\ \cline{2-4} %
         & GDC & Static & No GDC \\ \hline 
        1/4  & \textbf{-15.24} & -14.13 & -13.65 \\
        1/16 & \textbf{-9.81}  & -8.42  & -8.11  \\ 
        1/32 & \textbf{-6.88}  & -6.54  & -6.07  \\
        1/64 & \textbf{-4.70}  & -4.42  & -4.13  \\
        \Xhline{2\arrayrulewidth}
        \end{tabularx}
    \end{threeparttable}
\end{table}

The convergence analysis in Fig.~\ref{fig:ablation_study_SDC} (at $\eta=1/16$) further reinforces these findings. The GDC connections exhibit stable convergence, reaching approximately -9.81 dB. In contrast, the Baseline (Removed) model plateaus at a higher NMSE of -8.11 dB. Although the Static configuration converges quickly initially, its performance levels off at -8.42 dB, remaining 1.39 dB inferior to GDC.
These results collectively underscore the effectiveness of the proposed GDC structure. The combination of learnable gating and additive merging provides a significant advantage over both static dense connections and the absence of skip connections for CSI compression and reconstruction tasks.

\section{Simulation Results and Analysis}
\label{sect:results}

In this section, simulations are conducted to evaluate the performance of the proposed AiANet model from four key perspectives: (1) intra-scenario compression efficiency compared to state-of-the-art benchmarks, (2) cross-scenario generalization capability under separate-training, (3) generalization performance under mixed-training, and (4) computational complexity comparison. The evaluation is based on publicly available benchmark datasets \cite{wen2018deep}, with clearly defined experimental settings to ensure reproducibility.

\subsection{Simulation Setup}

The model was implemented using PyTorch 2.0 and trained with a batch size of 200. Weight matrices were initialized using the Xavier normal distribution \cite{glorot2010understanding} to mitigate potential gradient vanishing or explosion issues. We employed the AdamW optimizer \cite{loshchilov2019decoupled} with hyperparameters $\beta_1 = 0.9$, $\beta_2 = 0.999$, an initial learning rate $\eta_0 = 10^{-4}$, and weight decay disabled ($\lambda = 0$).
A cosine annealing scheduler \cite{loshchilov2017sgdr} dynamically adjusted the learning rate during training according to:
\begin{equation}\label{eq:cosine_annealing}
\eta_t = \eta_{\text{min}} + \frac{1}{2}(\eta_0 - \eta_{\text{min}})\left(1 + \cos\left(\frac{t}{T_{\text{max}}}\pi\right)\right),
\end{equation}
where $\eta_0 = 10^{-4}$ is the initial learning rate, $\eta_{\text{min}} = 10^{-6}$ is the minimum learning rate, $T_{\text{max}}=1000$ is the total number of training epochs, and $t \in [0, T_{\text{max}})$ denotes the current epoch index.
Parametric Rectified Linear Units (PReLU) \cite{he2015delving} with channel-wise learnable slopes were used as activation functions. PReLU helps preserve negative gradient flow while enhancing nonlinear representation capacity, which is particularly beneficial for encoding sparse CSI features.

\subsection{Performance Evaluation}
\subsubsection{Intra-Scenario Performance}

We first evaluate AiANet in intra-scenario settings, where models are trained and tested using data from the same channel environment (either indoor or outdoor). Performance, measured by NMSE in dB and cosine similarity ($\rho$), is compared against several benchmarks in Table~\ref{tb:AiANet_trained_on_single_type}.
\begin{itemize}
    \item \textbf{Indoor Performance:} AiANet consistently outperforms all benchmarks in the indoor scenario. Compared to the second-best performing model, ACRNet, AiANet achieves an average NMSE improvement of 2.22 dB across all compression ratios. 
    The largest gain over ACRNet is 3.42 dB at $\eta = 1/4$. 
    \item \textbf{Outdoor Performance:} AiANet also demonstrates superior performance in the more challenging outdoor scenario, maintaining an average NMSE advantage of 1.24 dB over ACRNet across all compression ratios. 
\end{itemize}

\begin{table}[htbp]
    \centering
    \footnotesize
    \caption{Intra-Scenario Performance Comparison (NMSE in dB and Cosine Similarity $\rho$).} 
    \label{tb:AiANet_trained_on_single_type}
    \renewcommand{\arraystretch}{1.3}
    \begin{tabularx}{\columnwidth}{|>{\centering\arraybackslash}X|l|>{\centering\arraybackslash}X>{\centering\arraybackslash}X|>{\centering\arraybackslash}X>{\centering\arraybackslash}X|} 
        \Xhline{2\arrayrulewidth}
        \multirow{2}{*}{$\eta$} & \multirow{2}{*}{Method}  & \multicolumn{2}{c|}{Indoor} & \multicolumn{2}{c|}{Outdoor} \\ \cline{3-6} 
         &  & NMSE  & $\rho$ & NMSE & $\rho$ \\ \hline
        \multirow{6}{*}{1/4}
        & CsiNet~\cite{wen2018deep}      & -17.36  & 0.99 & -8.75  &  0.91 \\
        & CsiNetPlus~\cite{Guo2020csinetplus}  & -27.37   &  1.00 & -12.40  & 0.96 \\
        & CRNet~\cite{lu2020multi}       & -26.99   & 1.00 & -12.70  &  0.96\\
        & DCRNet~\cite{tang2022dilated}  & -30.61 & 1.00 &  -13.61 &   0.96\\
        & ACRNet~\cite{lu2022binarized}  & -32.02 & 1.00 & -14.25  &   0.97\\
        & AiANet (proposed)  & \textbf{-35.44} & \textbf{1.00} & \textbf{-17.53} & \textbf{0.97} \\ \hline 
        \multirow{6}{*}{1/16}
        & CsiNet~\cite{wen2018deep}      & -8.65  & 0.93 & -4.51  & 0.79 \\
        & CsiNetPlus~\cite{Guo2020csinetplus} & -14.14   & 0.97& -5.73  &  0.86 \\
        & CRNet~\cite{lu2020multi}       & -11.35   & 0.95 & -5.44  & 0.86 \\
        & DCRNet~\cite{tang2022dilated}  & -14.26 &  0.97&  -6.35 &  0.89\\
        & ACRNet~\cite{lu2022binarized}  & -15.05 &  0.97& -6.47  &   0.90\\
        & AiANet (proposed)  & \textbf{-17.46}  & \textbf{0.98} & \textbf{-7.42}  & \textbf{0.90}  \\ \hline
        \multirow{6}{*}{1/32}
        & CsiNet~\cite{wen2018deep}      & -6.24  & 0.89 & -2.81  &  0.67 \\
        & CsiNetPlus~\cite{Guo2020csinetplus}& -10.43   &  0.95& -3.40  &  0.70 \\
        & CRNet~\cite{lu2020multi}       & -8.93   & 0.94 & -3.51  &  0.70 \\
        & DCRNet~\cite{tang2022dilated}  & -9.88 & 0.94 &-3.95  &   0.72\\
        & ACRNet~\cite{lu2022binarized}  & -10.77  & 0.95 & -4.05  &  0.73\\
        & AiANet (proposed)  & \textbf{-12.81}  & \textbf{0.96} & \textbf{-4.57}  & \textbf{0.79}  \\ \hline
        \multirow{6}{*}{1/64}
        & CsiNet~\cite{wen2018deep}      & -5.84 & 0.87 & -1.93 &  0.59 \\
        & CsiNetPlus~\cite{Guo2020csinetplus} & -6.14   &0.87  & -2.13  & 0.60  \\
        & CRNet~\cite{lu2020multi}       & -6.94   & 0.88 & -2.22  & 0.60 \\
        & DCRNet~\cite{tang2022dilated}  & -7.51 & 0.90 &  -2.44 &   0.61\\
        & ACRNet~\cite{lu2022binarized}  & -7.78  & 0.90 & -2.69  &  0.62\\
        & AiANet (proposed)  & \textbf{-8.69}  & \textbf{0.91} & \textbf{-2.90}  & \textbf{0.64}  \\
        \Xhline{2\arrayrulewidth}
    \end{tabularx}
\end{table}

\subsubsection{Cross-Scenario Generalization Test} \label{subsubsect:cross-scenario testing}

The cross-scenario generalization test involves training models exclusively on data from one scenario (e.g., indoor) and evaluating them on data from the other (e.g., outdoor). This procedure measures adaptability to varying channel conditions, as detailed in Table~\ref{tb:cross_scenario_performance}. 

Table~\ref{tb:cross_scenario_performance} reveals that direct cross-scenario application poses a significant challenge for all evaluated models, as indicated by generally high NMSE values and low cosine similarities ($\rho$). This underscores the difficulty models face when encountering CSI patterns substantially different from their training distribution.
Despite this challenge, the proposed AiANet demonstrates notably better cross-scenario generalization compared to the baseline methods, consistently achieving the lowest NMSE values across the evaluated compression ratios. 
For instance, at $\eta = 1/4$, AiANet achieves an NMSE of -0.27 dB and $\rho$ of 0.53 for the Indoor-to-Outdoor task, significantly outperforming the other models.
It is also observed that as compression rate increases (i.e., $\eta$ decreases), the performance advantage of AiANet over the other models (especially ACRNet and DCRNet) narrows remarkably. This suggests that at higher compression ratios, the severely limited information within the codeword hinders effective reconstruction for all models, diminishing the relative gains achieved by AiANet with less compression.

In summary, the cross-scenario tests highlight the inherent difficulty in directly applying CSI feedback models trained in one environment to another without adaptation. While the proposed AiANet exhibits superior robustness and generalization capability compared to existing benchmarks, especially at lower compression factors, the significant performance drop across all models emphasizes the domain gap between indoor and outdoor scenarios. To bridge this gap, more effective generalization strategies are needed for developing more universally applicable CSI compressors across scenarios.

\begin{table}[htbp]
    \centering
    \footnotesize
    \caption{Cross-Scenario Generalization Performance (NMSE in dB and Cosine Similarity $\rho$).}
    \label{tb:cross_scenario_performance}
    \renewcommand{\arraystretch}{1.3}
    \begin{tabularx}{\columnwidth}{|>{\centering\arraybackslash}X|l|>{\centering\arraybackslash}X>{\centering\arraybackslash}X|>{\centering\arraybackslash}X>{\centering\arraybackslash}X|} 
        \Xhline{2\arrayrulewidth}
        \multirow{2}{*}{$\eta$} & \multirow{2}{*}{Method}  & \multicolumn{2}{c|}{Indoor-to-Outdoor} & \multicolumn{2}{c|}{Outdoor-to-Indoor} \\ \cline{3-6} 
         &  & NMSE  & $\rho$ & NMSE  & $\rho$ \\ \hline
        \multirow{6}{*}{1/4}
        & CsiNet~\cite{wen2018deep}      & 8.13  & 0.19 & 10.22  & 0.18  \\
        & CsiNetPlus~\cite{Guo2020csinetplus}  & 6.24   & 0.21 & 8.20  & 0.19  \\
        & CRNet~\cite{lu2020multi}       & 7.88   & 0.20 & 9.03  & 0.19  \\
        & DCRNet~\cite{tang2022dilated}   & 4.06 & 0.27 & 5.50  & 0.24  \\
        & ACRNet~\cite{lu2022binarized}  & 1.55 & 0.36 & 3.22 & 0.28\\
        & AiANet (proposed)  & \textbf{-0.27} & \textbf{0.53} & \textbf{1.77}  & \textbf{0.39}  \\ \hline
        \multirow{6}{*}{1/16}
        & CsiNet~\cite{wen2018deep}      & 10.82  & 0.17 & 12.61  & 0.16  \\
        & CsiNetPlus~\cite{Guo2020csinetplus} & 8.89   & 0.18 & 10.53  & 0.17  \\
        & CRNet~\cite{lu2020multi}       & 9.72   & 0.17 & 11.14  & 0.16  \\
        & DCRNet~\cite{tang2022dilated}  & 6.54 & 0.20 & 8.96  & 0.18  \\
        & ACRNet~\cite{lu2022binarized}  & 5.81 & 0.22 & 7.94 & 0.19 \\ 
        & AiANet (proposed)  & \textbf{3.22} & \textbf{0.27} & \textbf{5.16}  & \textbf{0.22}  \\ \hline
        \multirow{6}{*}{1/32}
        & CsiNet~\cite{wen2018deep}      & 12.49  & 0.16 & 14.34  & 0.15  \\
        & CsiNetPlus~\cite{Guo2020csinetplus}& 10.07   & 0.18 & 13.41  & 0.16  \\
        & CRNet~\cite{lu2020multi}       & 10.95   & 0.18 & 14.83  & 0.15  \\
        & DCRNet~\cite{tang2022dilated} & 8.42 & 0.19 & 12.38  & 0.16\\
        & ACRNet~\cite{lu2022binarized}  & 8.38  & 0.19 & 11.72  & 0.17  \\
        & AiANet (proposed)  & \textbf{8.03} & \textbf{0.19} & \textbf{11.23}  & \textbf{0.17}  \\ \hline 
        \multirow{6}{*}{1/64}
        & CsiNet~\cite{wen2018deep}      & 15.36 & 0.14 & 16.46 & 0.14  \\
        & CsiNetPlus~\cite{Guo2020csinetplus} & 13.21   & 0.16 & 15.73  & 0.15  \\
        & CRNet~\cite{lu2020multi}       & 13.68   & 0.16 & 15.03  & 0.15  \\
        & DCRNet~\cite{tang2022dilated} & 11.93 & 0.17 & 13.38 & 0.16  \\  
        & ACRNet~\cite{lu2022binarized}  & 11.63  & 0.17 & 13.36  & 0.16  \\
        & AiANet (proposed)  & \textbf{11.56} & 0.17 & \textbf{13.33}  & \textbf{0.16} \\ 
        \Xhline{2\arrayrulewidth}
    \end{tabularx}
\end{table}

\subsubsection{Mixed-Training Performance}

\begin{figure}[htbp]
\centering
\subfigure[Original.]{\includegraphics[width=0.48\textwidth]{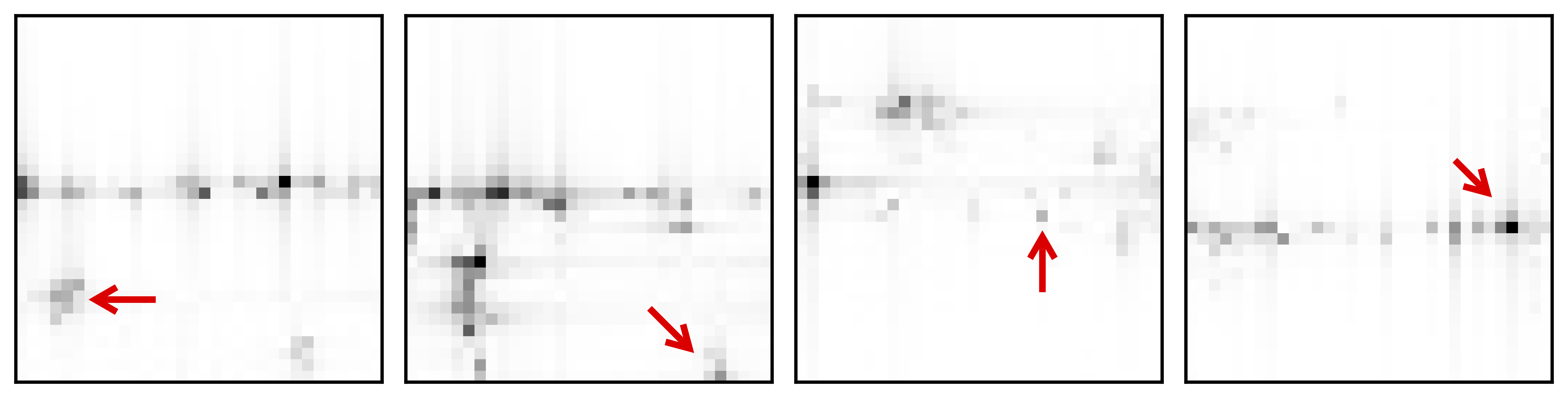}}
\subfigure[ACRNet reconstruction.]{\includegraphics[width=0.48\textwidth]{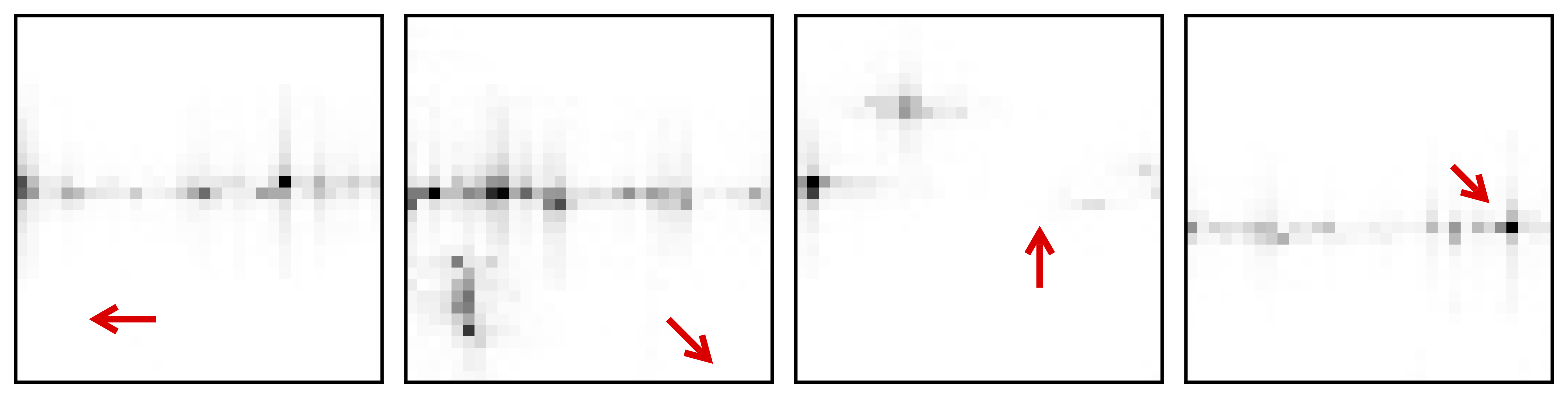}}
\subfigure[AiANet reconstruction.]{\includegraphics[width=0.48\textwidth]{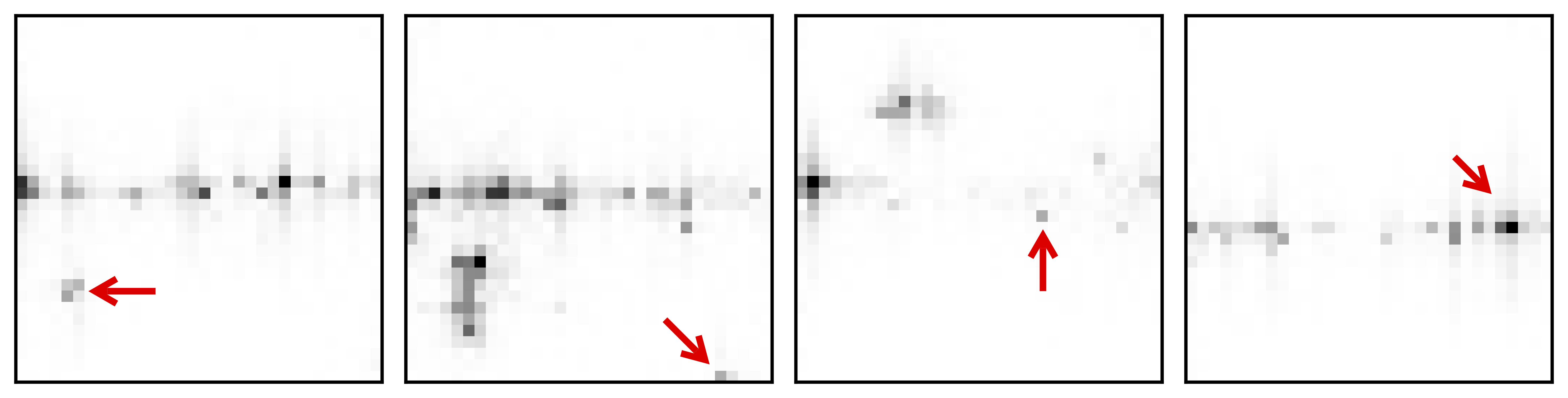}}
\caption{Visual comparison of CSI reconstruction under mixed-training when $\eta = 1/16$. 
Top row: original input; Middle row: reconstruction by ACRNet; Bottom row: reconstruction by the proposed AiANet. Red arrows indicate regions where differences are particularly evident. 
Compared to ACRNet, AiANet more accurately reconstructs fine-grained spatial features and sparse structures.}
\label{fig:restore_results}
\end{figure}

To further evaluate the generalization ability of AiANet, we assess its effectiveness in extracting CSI features from mixed CSI samples with disparate patterns --- specifically, those from indoor and outdoor channel scenarios.
We adopt a mixed-training strategy in which the model is trained on a balanced dataset composed of randomly interleaved indoor and outdoor samples. After training, the model is evaluated separately on pure indoor and pure outdoor test sets.
\begin{table*}[htbp]
    \centering
    \footnotesize 
    \caption{Performance Comparison of AiANet and ACRNet under Mixed-Training.}
    \label{tab:mixed-training_performance}
    \begin{tabular}{
        |c| 
        S[table-format=-2.2]  
        S[table-format=1.2]|  
        S[table-format=-2.2]  
        S[table-format=1.2]|  
        S[table-format=-2.2]  
        S[table-format=1.2]|  
        S[table-format=-2.2]  
        S[table-format=1.2]|  
    }
    \hline 
    \multirow{3}{*}{$\eta$} & \multicolumn{4}{c|}{Mixed-Trained-Indoor-Tested} & \multicolumn{4}{c|}{Mixed-Training-Outdoor-Tested} \\ 
    \cline{2-9} 
     & \multicolumn{2}{c|}{AiANet} & \multicolumn{2}{c|}{ACRNet} & \multicolumn{2}{c|}{AiANet} & \multicolumn{2}{c|}{ACRNet} \\ 
    \cline{2-9} 
    
     & {NMSE (dB)} & {$\rho$} & {NMSE (dB)} & {$\rho$} & {NMSE (dB)} & {$\rho$} & {NMSE (dB)} & {$\rho$} \\ 
    \hline 
    1/4  & -21.09 & 1.00 & -16.33 & 0.99 & -13.85 & 0.98 & -10.07 & 0.95 \\ 
    1/16 & -10.35 & 0.97 & -8.14 & 0.92 &  -6.87 & 0.89 & -5.75 & 0.85 \\ 
    1/32 &  -9.75 & 0.95 & -6.94 & 0.84 &  -4.91 & 0.81 & -4.12 & 0.78 \\ 
    1/64 &  -6.43 & 0.88 & -5.63  & 0.81 &  -2.81 & 0.68 & -2.45  & 0.63 \\
    \hline 
    \end{tabular}
\end{table*}
For direct comparison, we apply the same mixed-training approach to ACRNet, the strongest baseline in our benchmark, and evaluate its performance under identical conditions.
Visual reconstruction results from AiANet and ACRNet, compared to original samples, are shown in  Fig.~\ref{fig:restore_results}. Outdoor samples are selected as their higher noise levels make reconstruction more challenging, thereby providing clearer differences in reconstruction quality.
Comprehensive quantitative results across both indoor and outdoor datasets are detailed in Table~\ref{tab:mixed-training_performance}.

As Table~\ref{tab:mixed-training_performance} shows, mixed-training enables both AiANet and ACRNet to achieve much greater generalization capacity across scenarios compared with individual training and cross-scenario testing.
At $\eta = 1/4$, AiANet obtains NMSE value of -21.09 dB (indoor) and -13.85 dB (outdoor), indicating effective learning from the combined data and significantly outperforming ACRNet, which achieved -16.33 dB and -10.07 dB, respectively, under identical conditions. 
Performance degrades gracefully for both models with increasing compression, though AiANet consistently maintains lower NMSE values and higher $\rho$ correlation across all compression ratios.

Compared to intra-scenario training (Table~\ref{tb:AiANet_trained_on_single_type}), mixed-training involves a trade-off. For instance, at $\eta = 1/4$, the intra-scenario NMSE is significantly lower (-35.44 dB indoor, -17.53 dB outdoor) than the mixed-training results. This indicates that while mixed-training provides generalization, it sacrifices some degree of scenario-specific optimization.
A similar trade-off exists for ACRNet, although AiANet still achieves better absolute performance in the mixed-training setup.
However, in contrast to the poor generalization observed in the cross-scenario setting (Table~\ref{tb:cross_scenario_performance}), mixed training demonstrates genuine generalization capability, yielding far better performance. For example, at $\eta = 1/4$, mixed-trained AiANet achieves NMSEs of -13.85 dB on the outdoor test set and -21.09 dB on the indoor test set --- substantially outperforming its cross-scenario results discussed in Section~\ref{subsubsect:cross-scenario testing}. This indicates that investing in mixed-training, despite its trade-offs, is a worthwhile approach.
The improvement likely stems from the model learning universal features shared across CSI patterns from both scenarios, thereby avoiding the severe performance drop observed in direct cross-scenario transfer.
Notably, AiANet's mixed-training performance (-21.09 dB / -13.85 dB) is also substantially better than ACRNet's mixed-training performance (-16.33 dB / -10.07 dB), demonstrating its superior ability of capturing CSI features when trained on heterogeneous channel data.

\subsection{Computational Complexity Analysis}

We evaluate the computational complexity of AiANet using the number of trainable parameters and Floating-Point Operations (FLOPs), summarized in Table~\ref{tab:params_flops_comparison}. Analysis reveals three key aspects:

\begin{table*}[ht] 
    \centering
    \footnotesize
    \setlength{\tabcolsep}{4pt}
    \renewcommand{\arraystretch}{1.3}
    \caption{Comparison of Model Parameters and FLOPs across Different Architectures and Compression Ratios ($\eta$).}
    \label{tab:params_flops_comparison}
    \begin{tabular}{|c|c|c|c|c|c|c|c|} 
        \hline
        \multirow{2}{*}{Metric} & \multirow{2}{*}{$\eta$} & \multicolumn{6}{c|}{Network Architectures} \\ \cline{3-8} 
        & & AiANet (proposed) & CsiNet~\cite{wen2018deep} & CsiNetPlus~\cite{Guo2020csinetplus} & CRNet~\cite{lu2020multi} & ACRNet~\cite{lu2022binarized} & DCRNet~\cite{tang2022dilated} \\ \hline
        \multirow{4}{*}{Parameters}
        & $1/4$  & 2264\,K & 2103\,K & 2122\,K & 2103\,K & 2145\,K & 2115\,K \\ 
        & $1/16$  & 753\,K & 530\,K & 549\,K & 530\,K & 572\,K & 542\,K \\
        & $1/32$ & 408\,K & 268\,K & 286\,K & 257\,K & 309\,K & 279\,K \\
        & $1/64$ & 193\,K & 137\,K & 155\,K & 136\,K & 178\,K & 148\,K \\ \hline
        \multirow{4}{*}{FLOPs}
        & $1/4$  & 45.87\,M & 5.41\,M & 24.57\,M & 5.12\,M & 46.36\,M & 17.57\,M \\ 
        & $1/16$  & 43.62\,M & 3.84\,M & 23.01\,M & 3.55\,M & 44.79\,M & 16.00\,M \\
        & $1/32$ & 42.83\,M & 3.58\,M & 22.74\,M & 3.28\,M & 44.52\,M & 15.74\,M \\
        & $1/64$ & 42.64\,M & 3.45\,M & 22.61\,M & 3.16\,M & 44.39\,M & 15.23\,M \\ \hline 
    \end{tabular}
\end{table*}

\subsubsection{Parameter Count}
AiANet utilizes more parameters than the benchmark models across all compression ratios. For example, at $\eta = 1/4$, AiANet has approximately 7.7\% more parameters than CsiNet (2,264\,K vs. 2,103\,K). This is primarily due to its more complex architecture involving attention mechanisms. However, the number of parameters scales down significantly with increasing compression; the parameter count at $\eta=1/64$ is only about 8.5\% of that at $\eta=1/4$, showing efficient parameter reduction.

\subsubsection{Computational Load (FLOPs)}
AiANet exhibits a substantially higher FLOP count compared to simpler models like CsiNet and CRNet, requiring approximately 8.5$\times$ more FLOPs than CsiNet at $\eta = 1/4$ (45.87\,M vs. 5.41\,M). 
This increased computation stems from its deeper structure, dual attention mechanisms, and gated dense connections. Nevertheless, AiANet's FLOP count is comparable to that of ACRNet (45.87\,M vs. 46.36\,M at $\eta=1/4$), indicating competitive computational efficiency relative to other high-performance architectures.

\subsubsection{FLOPs Stability under Compression}

A notable characteristic is AiANet's relative stability in FLOPs across different compression ratios. While CsiNet's FLOPs decrease by 36\% from $\eta = 1/4$ to $\eta = 1/64$, AiANet's FLOPs decrease by only 7.0\%. This suggests that the core computational graph of AiANet remains largely active even under high compression, potentially contributing to its robust performance by preserving feature extraction capabilities.

In summary, AiANet achieves state-of-the-art performance and generalization at the cost of increased model parameters and computational complexity compared to simpler baselines, though its FLOPs are comparable to other advanced networks like ACRNet. The choice between AiANet and lighter alternatives depends on the specific application requirements, balancing performance needs against available computational resources and deployment constraints.

\section{Conclusion}
\label{sect:conclusion}

In this paper a novel learning-based CSI compression method named AiANet was proposed for massive MIMO systems. While the existing methods are all designed and trained for individual channel scenarios, the proposed model can handle CSI compression across indoor and outdoor scenarios, harnessing their discrepant channel characteristics. Specifically, through combining the adaptive attention fusion, localized self-attention mechanism, and learnable gated feature reuse, AiANet is capable of learning diverse CSI representations between different scenarios.  
Simulation results demonstrate that AiANet significantly outperforms state-of-the-art benchmarks, such as ACRNet, in reconstruction accuracy across various compression ratios, while maintaining comparable computational complexity. This enhanced accuracy provides a foundation for more precise downlink precoding, directly contributing to higher achievable data rates and improved spectral efficiency in massive MIMO systems.
Moreover, the proposed mixed-training scheme is shown to substantially improve cross-scenario generalizability. 
In summary, the proposed method has great potential to eliminate the need for environment-specific retraining and facilitate the practicability of massive MIMO deployment.

\bibliographystyle{IEEEtran}
\bibliography{IEEEabrv, Reference}

\end{document}